\documentclass[12pt]{article} 
\usepackage{amsmath}
\usepackage{amssymb}
\usepackage{amsthm}
\usepackage{tikz}
\usepackage{color}
\usepackage{array}
\usepackage{multirow}
\usepackage{url}
\usepackage{tabularx}
\usepackage{bm}
\usepackage{graphicx}
\usepackage{subfig}
\usepackage{hyperref}
\usetikzlibrary{arrows}
\usepackage{mdframed}
\usepackage{geometry}

\usepackage[font=footnotesize]{caption}

\newtheorem{theorem}{Theorem}[section]

\newcommand{\footremember}[2]{%
	\footnote{#2}
	\newcounter{#1}
	\setcounter{#1}{\value{footnote}}%
}

\DeclareMathOperator{\argmin}{argmin}
\DeclareMathOperator{\supp}{supp}
\DeclareMathOperator{\unif}{Unif}

\newcommand{\PP}{\mathbb{P}}

\newcommand{\ma}{\mathcal{A}}
\newcommand{\mc}{\mathcal{C}}

\newcommand{\ee}{\bm{e}}
\newcommand{\bb}{\bm{b}}
\newcommand{\bbo}{\bm{b}_0}

\newcommand{\be}{\bm{\eta}}

\newcommand{\xxi}{\bm{\xi}}
\newcommand{\va}{\bm{a}}

\newcommand{\vx}{\bm{x}}

\newcommand{\vv}{\bm{v}}

\newcommand{\la}{\langle}
\newcommand{\ra}{\rangle}

\newcommand{\lar}{\leftarrow}

\title{Thresholding Greedy Pursuit for Sparse Recovery Problems}
\author{Hai Le\footremember{alley}{Corresponding Author, hvl2@psu.edu}, Alexei Novikov\footremember{trailer}{anovikov@math.psu.edu }}
\date{}

\begin{document}
	\maketitle
	\abstract{We study here sparse recovery problems in the presence of additive noise. We analyze a thresholding version of the CoSaMP algorithm, named Thresholding Greedy Pursuit (TGP). We demonstrate that an appropriate choice of thresholding parameter, even without the knowledge of sparsity level of the signal and strength of the noise, can result in exact recovery with no false discoveries as the dimension of the data increases to infinity.}
		\tableofcontents


%
%
%
%

	\section{Introduction} \label{section1}

In this section, we introduce our algorithm and associated theorems for theoretical guarantees. We also give an overview of sparse recovery algorithms and related literatures.
\subsection{Sparsity Promoting Optimization}

We are interested in finding sparse  \textit{signals} $\vx \in \mathbb C^K$ from   \textit{measurements} $\bb \in \mathbb C^{N}$ that are related by
\begin{equation}\label{eq:sparse}
\ma \vx +\ee  = \bb,
\end{equation}
where  $\ma\in \mathbb C^{N\times K}$ is  the \textit{measurement matrix} and $\ee$ is an unknown noise vector. 
Typically, the system \eqref{eq:sparse} is \textit{underdetermined} because we can only gather a few measurements, so $N \ll K$. 
When $N \ll K$  it not possible to solve this system uniquely without additional a priori information. Then it is usually assumed that the 
signal vector $\vx$ is $M$-sparse, which means it has $M$ nonzero entries,  and $M \ll K$. In the noiseless case ($\ee =0 $), one can find a solution using the Basis Pursuit~\cite{Claerbout1973}:
\begin{equation}\label{eq:basispursuit}
\text{Find } \argmin\|\vx\|_1\quad \text{s.t. }\ma\vx = \bb. 
\end{equation}
The solution of the optimization problem~\eqref{eq:basispursuit} recovers the original signal $\vx$ exactly under certain conditions on the measurement matrix $\ma$ and sparsity level 
$M$~\cite{ Candes2006, Chen2001, Elad2002, Feuer2003}.

In the presence of noise, the exact recovery is no longer possible. However, again under certain assumptions on $\ma$  and $\vx$ we still can recover the support of $\vx$ with 
only the knowledge of $\ma$ and $\bb$. The idea is that since $\vx$ is sparse, only a few columns of $\ma$ are used to produce $\bb$. Then    we can effectively detect which columns are best to approximate $\bb$. 
A popular approach is a modification of \eqref{eq:basispursuit}, the Basis Pursuit Denoising (BPDN)~\cite{Chen1994}  or  the Least Absolute Shrinkage and Selection Operator (LASSO)~\cite{Tibshirani1996}. 
It  is an $l_2$-optimization method that promotes sparsity by penalizing $l_1$-norm:
\begin{equation}\label{eq:lasso}
\text{Find }\vx_{\lambda}\in \argmin\left(\frac{1}{2}\|\bb-\ma\vx\|_2^2+\lambda\|\vx\|_1 \right),\quad \lambda \ge 0.
\end{equation}
Convexity ensures that there is 
always a solution to~\eqref{eq:lasso}. The tuning/penalty parameter $\lambda$ is appropriately chosen to obtain the desired properties of the minimizer $\vx_{\lambda}$. 
As $\lambda$ increases, BPD will choose the minimizer with fewer non-zero entries. On the other hand, smaller $\lambda$ will produce a solution closer to the least-squares 
approximation.  Given the level of noise one can choose $\lambda$ optimally so that the support of $\vx_{\lambda}$   recovers as much of the  support of the true solution 
$\vx$ as possible, see for example, \cite{Fuchs2005,Tropp2006,Wainwright2009}. These conditions depend on the knowledge of variance 
of the noise $\ee$ and also on the choice of $\lambda$.  Some examples  to choose $\lambda$ include \textit{cross-validation} \cite{Tibshirani1996} and choosing $\lambda$ \textit{adaptively} \cite{Zou2006, Chichignoud2016}.
These methods may be  computationally expensive and/or  require additional estimates of, for example, the level of noise.
Therefore, there has been  some research on designing algorithms with parameters that are independent of the level of noise. These algorithms may require other constraints; for example, the work of~\cite{Laska2009} assumes the noise to be sparse, 
or the work of~\cite{Kueng2018} requires non-negativity of the signal. 
We  discuss here two algorithms that do not have such restrictions and are directly related to our work.
The first  work is \textit{Square-Root LASSO} \cite{BELLONI2011} which is the following optimization problem
\begin{equation}\label{eq:sqrtlasso}
\text{Find }\vx_{\lambda}\in \argmin\left(\|\bb-\ma\vx\|_2+\lambda\|\vx\|_1 \right),\quad \lambda \ge 0.
\end{equation}
The functionals in~\eqref{eq:sqrtlasso}  and in~\eqref{eq:lasso}
differ in the exponent of the $l_2$-norm. This difference allows to choose $\lambda$ independent of knowledge of noise in the Square-Root LASSO approach~\cite{BELLONI2011}.
The authors called their method pivotal with respect to $\lambda$,  because they  do not need to know the level of noise to choose $\lambda$.  Our algorithm is a greedy implementation
of~\eqref{eq:sqrtlasso}. 
The second method is the \textit{Noise Collector}~\cite{Moscoso2020,Moscoso2020a}.  The Noise Collector is the Basis pursuit  applied to the following augmented linear system
\begin{equation}\label{eq:nc}
\text{Find $(\vx_{\tau},\be_{\tau})\in \argmin (\tau\|\vx\|_1+\|\be\|_1)$, subject to $\ma \vx+\mc \be = \bb$,}
\end{equation}
where  $\mc$ is  the noise  collector matrix, and $\tau$ is a tuning parameter.  
If the columns of $\mc$ are drawn at random and the parameter $\tau$ is chosen appropriately, then  the noise collector will ``absorb'' \textit{all} the noise:  
$\mc \be \approx \ee$ (and some signal) for any level of noise~\cite{Moscoso2020a}. Further
the Noise Collector method is also pivotal with respect to $\tau$.  Our proofs are inspired by the proofs for the Noise Collector by~\cite{Moscoso2020a}.  

The objective of this paper is to  
propose a \textbf{Thresholding Greedy Pursuit} (TGP), a fast algorithm  that finds solutions of~\eqref{eq:sqrtlasso}.
\begin{figure*}[ht] \centering 
	\captionsetup{width=.8\linewidth}
	\subfloat{\includegraphics[width=0.5\textwidth]{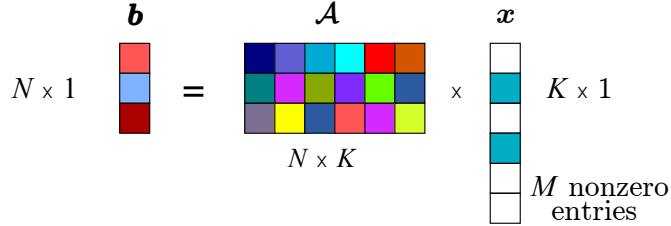}} 
	\caption{An illustation for Sparse Recovery Problems: The system is underdetermined as $\ma$ has more columns than rows; $\vx$ has several non-zero (blue) entries, while $\ma$ and $\bb$ are known and random in general.}
\end{figure*}
Similar to Noise Collector and Square-Root LASSO, TGP does not have parameters 
that depend on the level of noise.  
TGP is a \textbf{Greedy Pursuit}.  Greedy Pursuits are an important category of sparse recovery algorithms. 
Their idea is to build up an approximation one step at a time by making locally optimal choices at each step.
The representatives are Orthogonal Matching Pursuit (OMP)~\cite{Tropp2007} and  Compressive Sampling Matching Pursuit (CoSaMP)~\cite{Needell2009}.
The advantages of these algorithms are speed and sampling efficiency.  
TGP is a modification of  CoSaMP.  The first difference between  CoSaMP and TGP 
lies in the greedy selection criteria: CoSaMP chooses the largest entries of the proxy signal, TGP chooses all the entries that are above a certain threshold.
The second difference is in the choice of parameters.  CoSaMP needs to know the sparsity level to perform optimally.  TGP does not need it.

The paper is organized as follows. In Section \ref{section1}, we introduce our main results regarding the Thresholding Greedy Pursuit; we also give an overview of other sparse recovery algorithms. In Section \ref{section2}, we explain how the algorithm works and compare performance with its predecessor, CoSaMP. Finally, in Section \ref{section3}, we prove theoretical results about TGP and discuss future directions.

\subsection{Notations}
For any tall full-rank matrix $A$, denote its pseudoinverse $A^{\dag}:=(A^*A)^{-1}A^*$. For any set of indices $S$, denote $\ma_S$ as the matrix with columns of $\ma$ drawn from the set $S$. 
Note that the matrix operator $\ma_{S}\ma_{S}^{\dag}$ represents the orthogonal projection onto the vector space spanned by columns of $\ma$ indexed by the set $S$. 
We also denote $\supp(\vx)$ as the set of indices of nonzero entries of a vector $\vx$, and call that the support of this vector.  
Each column of $\ma$ is normalized to have unit $l_2$-norm. 
Bold letters $\va,\bb,\ldots$ are reserved for column vectors.
The length of the signal is $K=N^{\gamma}$ for some $\gamma \ge 1$.
The signal vector $\vx$ is expressed as $\vx:= (x_1,\ldots,x_K)$. For any set of indices $S$, denote the restriction to $S$ of $\vx$ as $\vx_S:=(x_k)_{k\in S}$.
We use $\| \vx \|_p$ to denote the  $l_p$-norm of a vector $\vx$ and $\|A\|:=\max\limits_{\|\vx\|_2=1}\|A\vx\|_2$ to denote the operator norm of $A$.
The identity matrix is denoted by $I$ whose dimension can vary. For any $N>0$, we use $\unif(\mathbb S^{N-1})$ to denote the uniform distribution on the $N$-dimensional unit sphere. Denote $\la \bm{u},\bm{v}\ra:= \bar{\bm u}^T\bm v$ the complex inner product between two complex vectors $\bm{u}$ and $\bm{v}$. We also denote $\text{Re}(x)$ as the real part of a complex number $x$.

\newpage

\subsection{Thresholding Greedy Pursuit Algorithm}

\begin{mdframed}
	
	\textsf{Thresholding}\textsf{ Greedy Pursuit Algorithm}
	
	\textbf{INPUT: }\text{measurement matrix $\ma\in \mathbb C^{N\times K}$, measurement vector $\bb\in \mathbb C^N$}, thresholding parameter $\tau>0$.
	
	\textbf{OUTPUT: }\text{a set $\Omega$ of indices of columns of $\ma$}.
	\smallskip
	
	\quad$\vx^0 \lar 0, \bb^0 \lar \bb, \Omega^0 \lar \supp(\vx^0)$\hfill \textbf{(Initialization)}
	\smallskip
	
	\textbf{repeat }
	\smallskip
	
	\quad$\vx^{n+1} \lar \dfrac{\ma^* \bb^n}{\|\bb^n\|_2}$\hfill \textbf{(Proxy Signal)}
	
	\quad$\vx^{n+1} \lar \max(|\vx^{{n+1}}|-\tau,0)$\hfill \textbf{(Thresholding)}
	
	\quad$\text{If $\vx^{n+1}=0$, then \textbf{break};}$ \hfill\textbf{(Stopping Criterion 1)}
	
	\quad$\Omega^{n+1}\lar\Omega^n\cup\supp(\vx^{n+1})$ \hfill\textbf{(Support Merging)}
	
	\quad$\bb^{n+1}\lar \bb-\ma_{\Omega^{n+1}}\ma_{\Omega^{n+1}}^{\dag}\bb$ \hfill\textbf{(Complement Projection)}
	
	\quad$\text{If $\|\bb^{n+1}\|_2=0$, then \textbf{break};}$ \hfill\textbf{(Stopping Criterion 2)}
	\smallskip
	
	\textbf{until }\text{stopped}
	
\end{mdframed}

The Thresholding Greedy Pursuit Algorithm is an iterative algorithm that produces a sequence of sets of indices $\Omega^1,\Omega^2,\ldots,$ that try to match the support of the signal vector $\vx$ by projecting the measurement 
vector on vector spaces that are orthogonally complement to vector spaces spanned by some columns of $\ma$. The detection at each iteration is done by a proxy step and a thresholding procedure. The output
of the algorithm is a set of indices, which we denote $\Omega$. The role of each step is as follows.
\begin{itemize}
	\item \textbf{Initialization}: The first approximation to the support is the empty set.
	\item \textbf{Proxy Signal}: Sparse recovery algorithms often rely on the fact that columns of the measurement matrix are almost orthogonal or not too collinear. If that is the case, large entries the vector $\ma^* \bb=\ma^*(\ma \vx+\ee)$ will be likely to match large entries of the vector $\vx$. As a consequence, at each iteration, large entries of the vector $\vx^{n+1}=\ma^*\bb^n/\|\bb^n\|_2$ will be good candidates for where the true support lies. The next step will show how to retrieve these large entries.
	\item \textbf{Thresholding}: This is the heart of the algorithm. This procedure will remove all the small entries of the proxy signal that are below a certain threshold $\tau$. This thresholding parameter is chosen so that the algorithm will not produce any false discoveries (locations that are detected by the algorithm but not in the true support; in other words, those are the indices that are in $\Omega$ but not in $\supp(\vx)$).
	\item \textbf{Stopping Criterion 1}: The algorithm will stop if nothing new is detected in the thresholding step.  
	\item \textbf{Support Merging}: New index set $\Omega^{n+1}$ is created by merging the old index set $\Omega^{n}$ and new locations in $\supp(\vx^{n+1})$.
	\item \textbf{Complement Projection}: We need to remove the part of measurement vector $\bb$ that approximates the locations that are already detected. This is done by projecting $\bb$ onto the space that is orthogonally complement to the space spanned by columns in $\ma_{\Omega^{n+1}}$.
	\item \textbf{Stopping Criterion 2}: The algorithm will stop if the whole support is detected, or if $\bb^{n+1}=\bb-\ma_{\Omega^{n+1}}\ma_{\Omega^{n+1}}^{\dag}\bb=0$, and so $\Omega^{n+1}$ will be the exact support of $\vx$.
\end{itemize}
The recovered signal with the support $\Omega$ is $\ma_{\Omega}^{\dag}\bb$, which is the solution to the $l_2$-approximation problem $\min\limits_{\vx}\|\ma_{\Omega}\vx-\bb\|_2$. The value of $\tau$ plays an important role in the performance of the algorithm. The next sections will show how we choose this parameter, independent of knowing the strength of the noise $\ee$, in order to guarantee support detection with no false discoveries.
\subsection{Main  Theorems}\label{sec:maintheorem}

We are ready to state our main results.

\begin{theorem}\label{theorem:nophantom} (\textbf{No Phantom Signal}) Let $\ma\in \mathbb{C}^{N\times K}$, $K=N^{\gamma}$.
	Suppose  there is no signal, that is $\vx=0$ in~\eqref{eq:sparse}, and the noise  $\ee$ in~\eqref{eq:sparse} is such that  
	$\ee/\| \ee \|_{\ell_2}$ is uniformly distributed on   $\mathbb{S}^{N-1}$.   For any $\kappa>0$ there exists   $c_0=\sqrt{2(\gamma+\kappa)}$ such that for any  $\tau\ge c_0\sqrt{\log N}/\sqrt{N}$ 
	the set  $\Omega$, the output of the TGP algorithm, is empty with probability $1-2/N^{\kappa}$.  
\end{theorem}

The above theorem guarantees the algorithm does not recover anything if the input is pure noise. The bound $ \tau \geq c_0\sqrt{\log N}/\sqrt{N}$  comes from estimating how large the inner product of random vectors in high      
dimensions could be. We follow~\cite{Vershynin2018} in obtaining this and similar estimates. The next theorem assures zero False Discovery Rate even when there is some information in the signal, that is the algorithm does not detect any entry outside of the support of the signal.  In order to guarantee the zero False Discovery Rate we need to assume incoherence of columns of $\ma$. Define the mutual coherence parameter 
\begin{equation}\label{inco}
\mu:= \max\limits_{1\le i<j\le K}|\la \va_i,\va_j\ra|.
\end{equation}

\begin{theorem}(\textbf{No False Discoveries}) \label{theorem:nofalse}
	Assume $\ma$, $\gamma$, $\kappa$, $c_0$,  $\tau$, $\ee$, $\Omega$ are as in the previous theorem, and $\mu$ is given by~\eqref{inco}.  
	Suppose $\vx$ is the $M$-sparse solution of~\eqref{eq:sparse}.
	If    $M\le 1/(4\mu)$,
	then  $\Omega \subset\supp(\vx)$ with probability $1-2/N^{\kappa}$.
\end{theorem}
The next theorem shows that, if the noise is not large, our method recovers the exact support of the signal.
\begin{theorem} (\textbf{Exact Recovery})\label{theorem:exact}
	Assume  $\ma$, $\gamma$, $\kappa$, $c_0$, $\mu$, $\ee$, $\Omega$  are as in the previous theorem, but
	\begin{align}\label{mutualcondition}
	M\le\min\left\{\frac{1}{4\mu},\frac{\sqrt{N}}{4c_0\sqrt{\log N}} \right\},
	\end{align}
	and
	\begin{equation}\label{threspara}
	\tau:= \sqrt{\frac{4}{3}\left(\frac{\mu}{4}+\frac{c_0^2\log N}{N} \right)}.
	\end{equation}
	If	$\|\ee\|_2\le 0.03 \sqrt{M} \min_{i \in {{\scriptsize{ \supp}}}(\vx) } (|x_i|)$,	
	then $\Omega = {{{\supp}}}(\vx) $   with probability $1-2/N^{\kappa}$.   
\end{theorem}

{\it Remark.} In the last Theorem the pessimistic constant $0.03 \sqrt{M}$ could be improved, see  formulas~\eqref{eq:noisestrength} and~\eqref{eq:functionf}. We keep  $0.03 \sqrt{M}$ for simplicity of presentation.

\subsection{Overview of Sparse Recovery Algorithms and Our Contributions}
Sparse recovery problems have found applications in the fields of compressed sensing \cite{Eldar2009}, signal denoising \cite{Gupta2016}, optical imaging \cite{Dileep2020}, machine learning \cite{Yang}, and more. The key idea that drives sparse recovery is that a high dimensional sparse signal can be inferred from a few linear observations. An intensive survey of many sparse recovery algorithms is presented in \cite{CrespoMarques2019}. They typically fall into three categories: \textbf{Combinatorial Algorithms}, \textbf{Convex Relaxation}, and \textbf{Greedy Pursuits}.

The first category requires a large number of structured samples of the signal for reconstruction via group testing. This includes Fourier sampling \cite{Gilbert2002,Gilbert2003}, chaining pursuit \cite{Gilbert2006} and HHS pursuit \cite{Gilbert2007}. The algorithms in this group are extremely fast but demand a huge number of samples that are not easy to obtain. 

The second class, Convex Relaxation, mostly deals with the BPDN problem \cite{Chen1994,Tibshirani1996} defined by
\[
\min\left(\frac{1}{2}\|\bb-\ma\vx\|_2^2+\lambda\|\vx\|_1 \right).
\]
The solution to this minimization problem achieves two objectives at the same time: solving the linear algebra problem while maintaining a small $l_1$-norm. The techniques to solve the optimization problem include interior-point methods \cite{Candes2006}, projected gradient \cite{Figueiredo2007}, and iterative thresholding \cite{Daubechies2004}. These algorithms require a very small number of measurements, but they rely on choosing the parameter $\lambda$ carefully to achieve the best performance \cite{Homrighausen2018}.

Our algorithm falls into the last class of being ``greedy." Examples include OMP \cite{Tropp2007}, stagewise OMP \cite{Donoho2012}, regularized OMP \cite{Needell2009a}, and CoSaMP \cite{Needell2009}. These methods build up 
an approximation by making locally optimal choices at each iteration. Their advantages are being fast and requiring modest samplings. The idea of using greedy algorithms in signal processing was used by \cite{Mallat1993} who also coined the name \textit{matching pursuit} for one of them. \cite{Gilbert2002,Gilbert2002a} developed fast algorithms of greedy nature for sparse approximation and
established novel rigorous guarantees for greedy methods. \cite{Tropp2007} then proposed a greedy iterative algorithm called \textit{orthogonal matching pursuit} (OMP)  and proved the algorithm was effective for compressive sampling. OMP takes the form similar to our description of TGP above. While TGP uses thresholding, OMP chooses the location of the column vector of $\ma$ that makes the largest inner product of the form $|\la \va_i,\bb\ra|$. \cite{Needell2009} built upon OMP to create an algorithm called \textit{compressive sampling matching pursuit} (CoSaMP). Instead of choosing the largest component, CoSaMP identifies many large inner products at each iteration and create the support from those. The analysis of CoSaMP is based on an important feature of the measurement matrix called  the \textit{restricted isometry propery} (RIP). This property was introduced by \cite{Candes2006,Candes2006a} in their work on convex relation methods. 		
RIP quantifies how a matrix preserves the distance between signals and it is crucial for the measurement matrix to have such property to be able to recover sparse signals. RIP is also important in another version of OMP called \textit{regularized OMP} (ROMP) which was developed by \cite{Needell2010}. By improving OMP, the authors established that under RIP the algorithm can also work with noisy data. 

Convergence theory of TGP could be developed if one assumes a RIP condition instead of assuming smallness of $\mu$ in~\eqref{inco}.
We chose to work with the  incoherence condition \eqref{inco} because  RIP is computationally harder to check. 
Moreover, deterministic matrices satisfying RIP are difficult to construct. The best result in this direction is the work of~\cite{Bourgain2011}.  In addition, we are motivated by sparse recovery problems in imaging applications, 
where the measurement matrix may not satisfy RIP. 

Our Thresholding Greedy Pursuit algorithm was inspired by the Noise Collector algorithm~\cite{Moscoso2020,Moscoso2020a}. 
We realized the analysis of the Noise Collector can be applied to the Square-Root LASSO \cite{BELLONI2011} as well. 
Then,  Anna Gilbert suggested to look at CoSaMP and investigate whether its greedy framework also could be covered by the  analysis of the Noise Collector. 
The TGP algorithm is the result of these three ingredients. The TGP uses the conjugate gradient to update the measurement vector.
This type of update is not new. For example, this idea was used in an algorithm of~\cite{Donoho2012} called \textit{stagewise OMP} (StOMP). 
Their method of choosing the thresholding parameter $\tau$ is based on the assumption that columns of $\ma$ are normally distributed. 
To the best of our knowledge, no rigorous results are available for StOMP. 
The conjugate gradient update is also used by~\cite{Yang2015a} when columns of $\ma$ satisfy RIP. Their analysis requires the knowledge of the strength of the noise in choosing its parameter. 
On the other hand, we are interested in usage of sparse recovery problems in imaging applications, the measurement matrix may not have normally distributed columns or satisfy RIP. 
Our contributions are the three results above in which by carefully choosing the thresholding parameter we can rigorously guarantee exact recovery even with noisy data.

\section{Ideas of the Proofs and Performance of TGP} \label{section2}
In this section, we explain the main ideas to prove the main theorems. We then compare the performance of TGP and CoSaMP in various settings.
\subsection{An Outline of the Proofs}
This outline contains the ideas for all three theorems. For  simplicity of presentation, assume that we want to prove Theorem~\ref{theorem:exact}, and that 
the signal is nonzero only in its first entry. We then can write the measurement vector as 
\[
	\bb = x_1\va_1+\ee.
\]
The algorithm will detect $x_1$ in the first iteration if the following inequality holds:
\begin{equation}\label{geq}
\frac{|\la \va_1,\bb\ra|}{\|\bb\|_2} > \tau.
\end{equation}
However, we also must ensure that any other column  of $\ma$ is not detected. Thus we must have
\begin{equation}\label{leq}
\frac{|\la \va_{i},\bb\ra|}{\|\bb\|_2}\le \tau, 
\end{equation}
for all $i \neq 1$. 
The algorithm will perform correctly if we can choose $\tau$ so that  both inequalities~\eqref{geq} and~\eqref{leq} are true. The condition \eqref{mutualcondition} will play a vital role in estimating the right value of $\tau$. 

We have now found  $x_1$. In the next iteration we want to remove dependence on $\va_1$ by projecting the measurement vector onto the space that is orthogonal to the vector space spanned by $\va_1$. 
Hence, the new measurement vector that goes into the next iteration will be
\[
\bb^{1}:=  \bb_{\text{new}}= \bb - \la \va_1,\bb\ra \va_1 = x_1\va_1+\ee-\la \va_1,x_1\va_1+\ee\ra\va_1= \ee - \la \va_1,\ee\ra\va_1.
\]
Now, if condition~\eqref{leq}  holds for all $i$, then we stop.  

Note that the term $x_1\va_1$ has disappeared in the new measurement vector  $\bb^{1}$.  The remainder is essentially the noise vector $\ee$ as the expression $\la \va_1,\ee\ra\va_1$ can be 
made small in a high dimensional space. Therefore~\eqref{leq} implies that if
\begin{equation}\label{leq2}
\frac{|\la \va_{i},\ee \ra|}{\|\ee \|_2}\le \tau/2, \mbox{ for all }  i.
\end{equation}
then we can detect all nonzero entries of $\vx$.
We will then make use of the fact that $\ee/\|\ee\|_2$ is uniformly distributed on the unit sphere and 
therefore it is essentially Gaussian. This will give us a lower bound estimate on the level of noise that we can handle.

If there are more than one nonzero entry of $\vx$,   we will use induction to show that at each iteration at least one of the remaining nonzero $x_i$ 
is detected  by the algorithm. More specifically, we will prove that at the $n$-iteration the projected measurement vector $\bb^{n}:=  \bb_{\text{new}}$ will satisfy
\begin{equation}\label{eq:exact}
|\la \va_{i_n},\bb^n\ra|> \tau \|\bb^n\|_2
\end{equation}
for at least some $i_n$ with $x_{i_n} \neq0$.  This implies that we will always detect at least one new $x_{i} \neq0$. It also implies that  the algorithm will always stop after at most $M$ iterations. 
The details are provided in Section \ref{section3}.

\subsection{Comparison with CoSaMP}
In this section we compare TGP  with CoSaMP,  its predecessor \cite{Needell2009}. CoSaMP takes measurement matrix $\ma \in \mathbb C^{N\times K}$ and measurements $\bb\in \mathbb C^N$ as inputs. Each iteration has computational complexity of order $O(NK)$. In comparison, we detail the computational complexity of TGP at each iteration:
\begin{itemize}
	\item \textbf{Proxy Signal} step takes $NK$ flops to compute $\ma^*\bb^{n}$ and $O(N)$ flops to compute $\|\bb^n\|_2$.
	\item \textbf{Thresholding} step takes $K$ flops to sweep through $\vx^{n+1}$.
	
	\item \textbf{Support Merging} takes no more than $O(M)$ flops.
	\item Inside the \textbf{Complement Projection} step, Conjugate Gradient (CG) solver \cite{Golub2012} takes at most $\nu\cdot 2NK+O(K)$ flops where $\nu$ is a fixed number of iterations for CG, and then, one application of $\ma$ takes an additional $NK$ flops. We note that the condition $\max\limits_{i\neq j}|\la \va_i,\va_j\ra|\le 1/(4M)$ implies small condition number for the matrix $\ma_{\Omega}^*\ma_{\Omega}$ for any $|\Omega|\le M$, which is beneficial for CG solver:
	\[
	\kappa(\ma_{\Omega}^*\ma_{\Omega}) = \frac{\lambda_{\max}(\ma_{\Omega}^*\ma_{\Omega})}{\lambda_{\min}(\ma_{\Omega}^*\ma_{\Omega})}\le \frac{5/4}{3/4}= \frac{5}{3} \approx 1.667.
	\]
\end{itemize}
Overall, the operation count for each iteration of TGP amounts to $(2\nu+2)NK+O(K)$. We note that CoSaMP also requires the user to specify how many iterations it will run; CoSaMP needs to have an estimate on the sparsity level $M$ to be efficient. In contrast, TGP is fully automatic and it will surely stop after $M$ iterations, without even knowing $M$. Numerics indicates that TGP stops after at most 2 iterations and still achieves a good performance.

\begin{figure}[ht] \centering 
	\captionsetup{width=.8\linewidth}
	\subfloat{\includegraphics[width=0.333\textwidth]{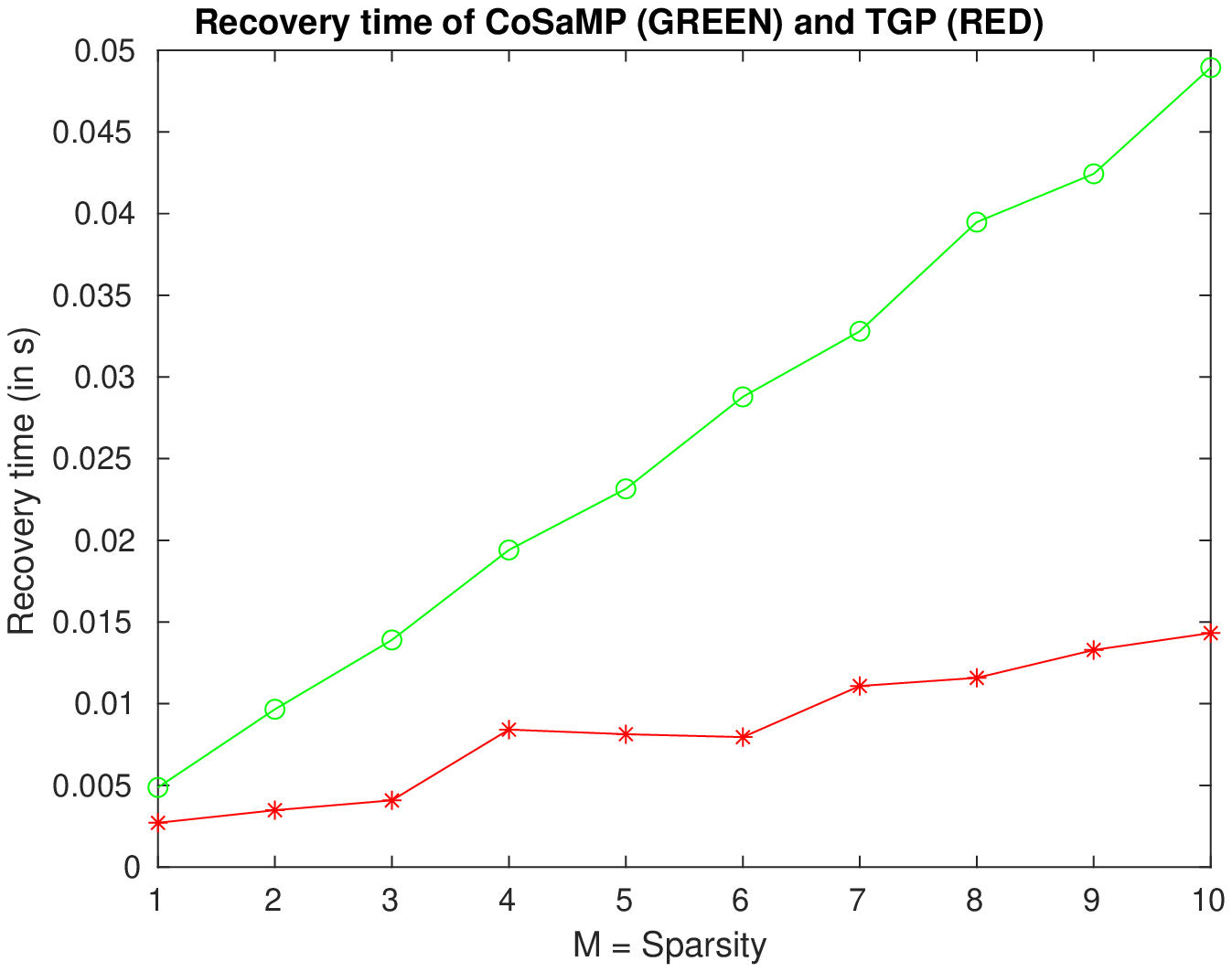}}  \subfloat{\includegraphics[width=0.333\textwidth]{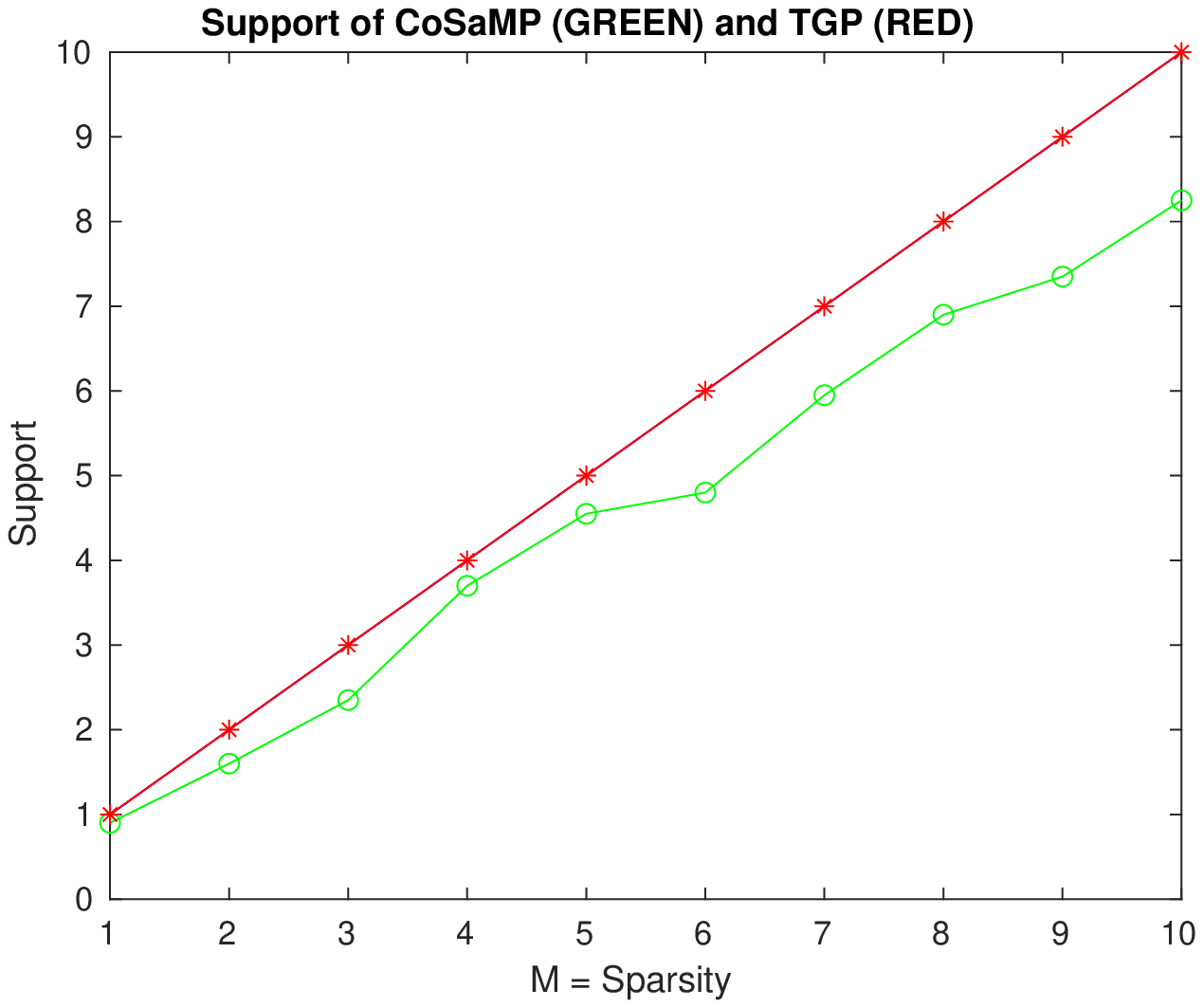}}
	\subfloat{\includegraphics[width=0.333\textwidth]{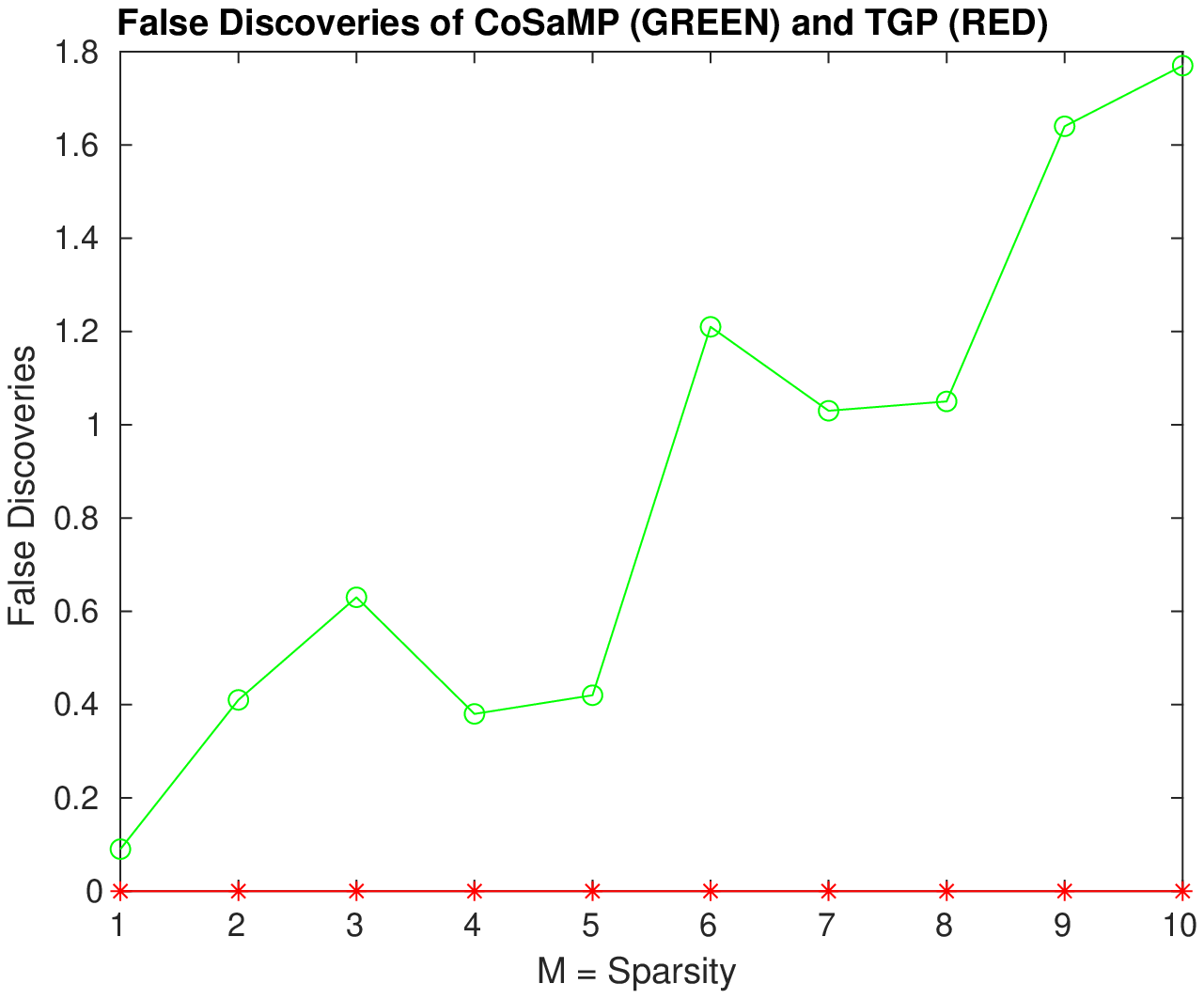}}
	\caption{$\ma$ is a Gaussian matrix, noise level is $\delta=0$.}
	\label{fig:gauss0}
\end{figure}

\begin{figure}[ht] \centering 
	\captionsetup{width=.8\linewidth}
	\subfloat{\includegraphics[width=0.333\textwidth]{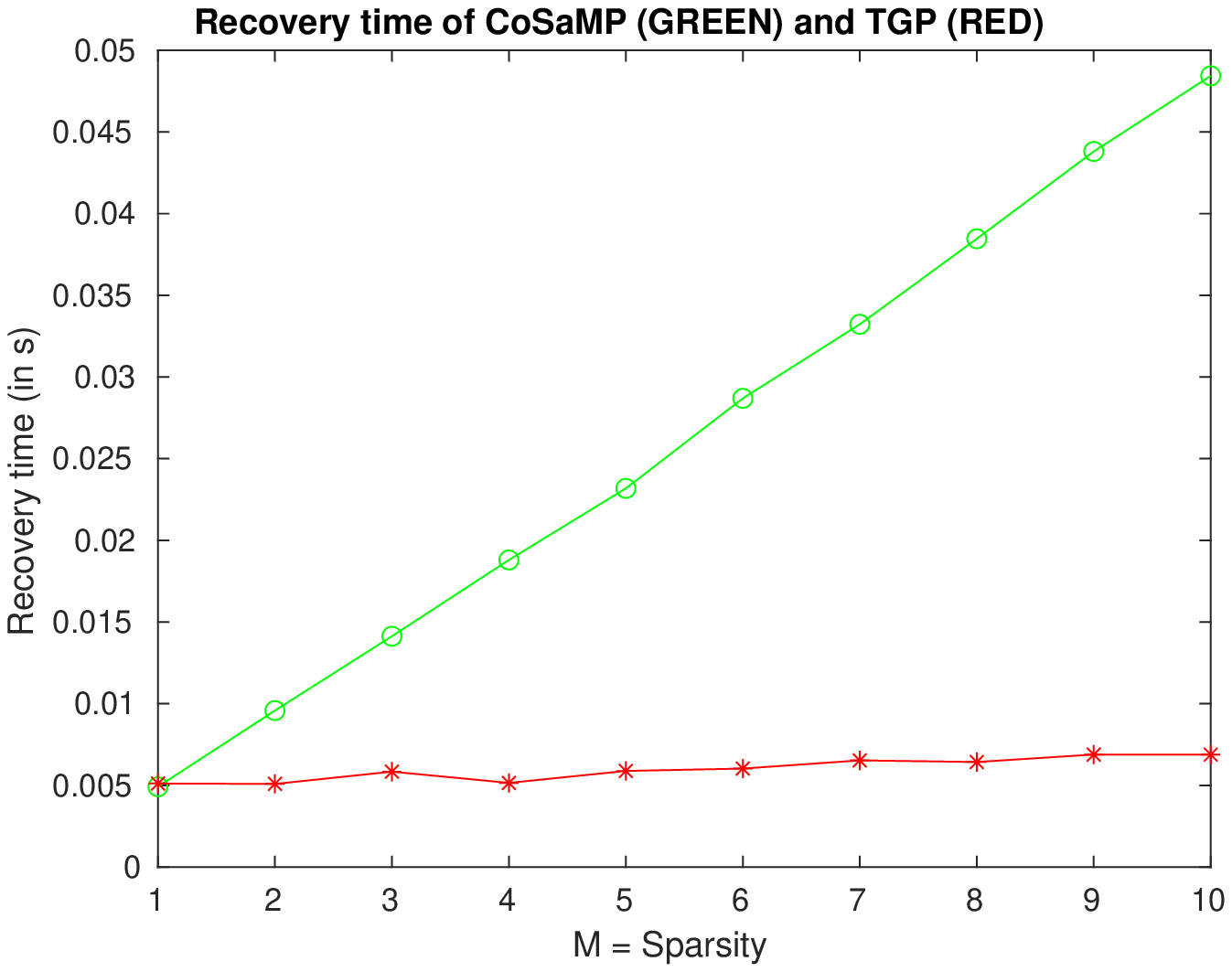}}  \subfloat{\includegraphics[width=0.333\textwidth]{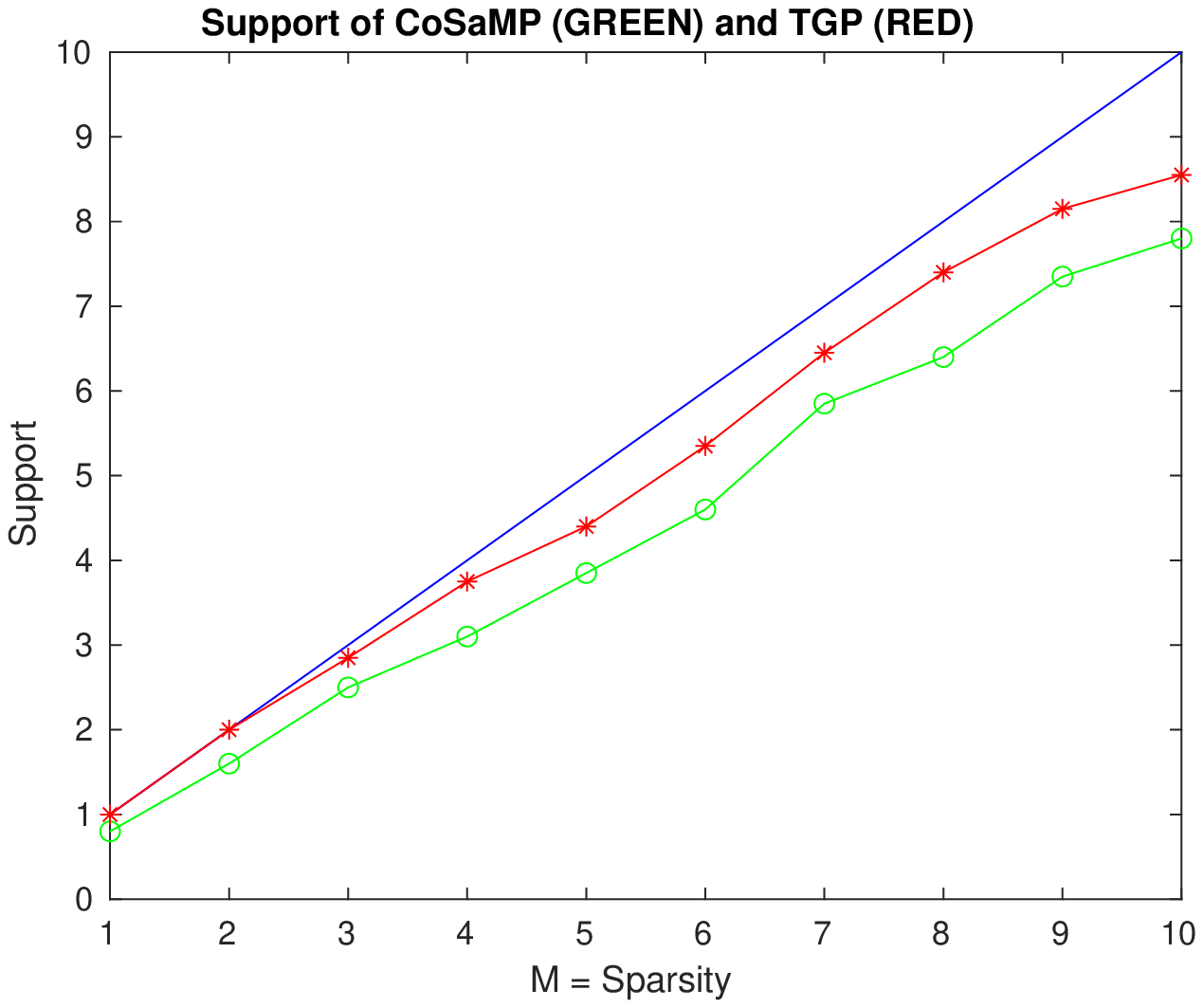}}
	\subfloat{\includegraphics[width=0.333\textwidth]{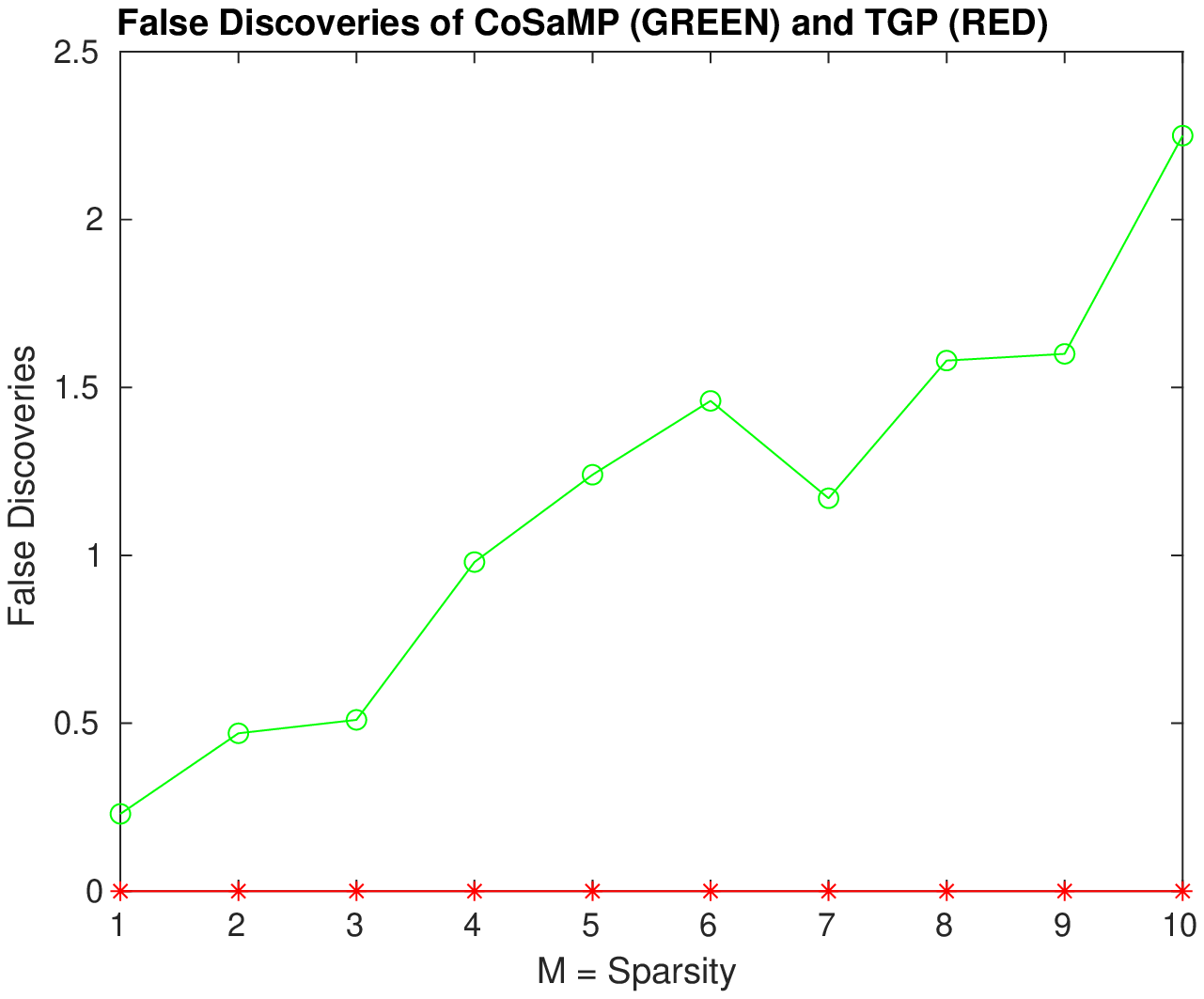}}
	\caption{$\ma$ is a Gaussian matrix, noise level is $\delta=0.5$.  }
	\label{fig:gauss05}
\end{figure}

\begin{figure}[ht] \centering 
	\captionsetup{width=.8\linewidth}
	\subfloat{\includegraphics[width=0.333\textwidth]{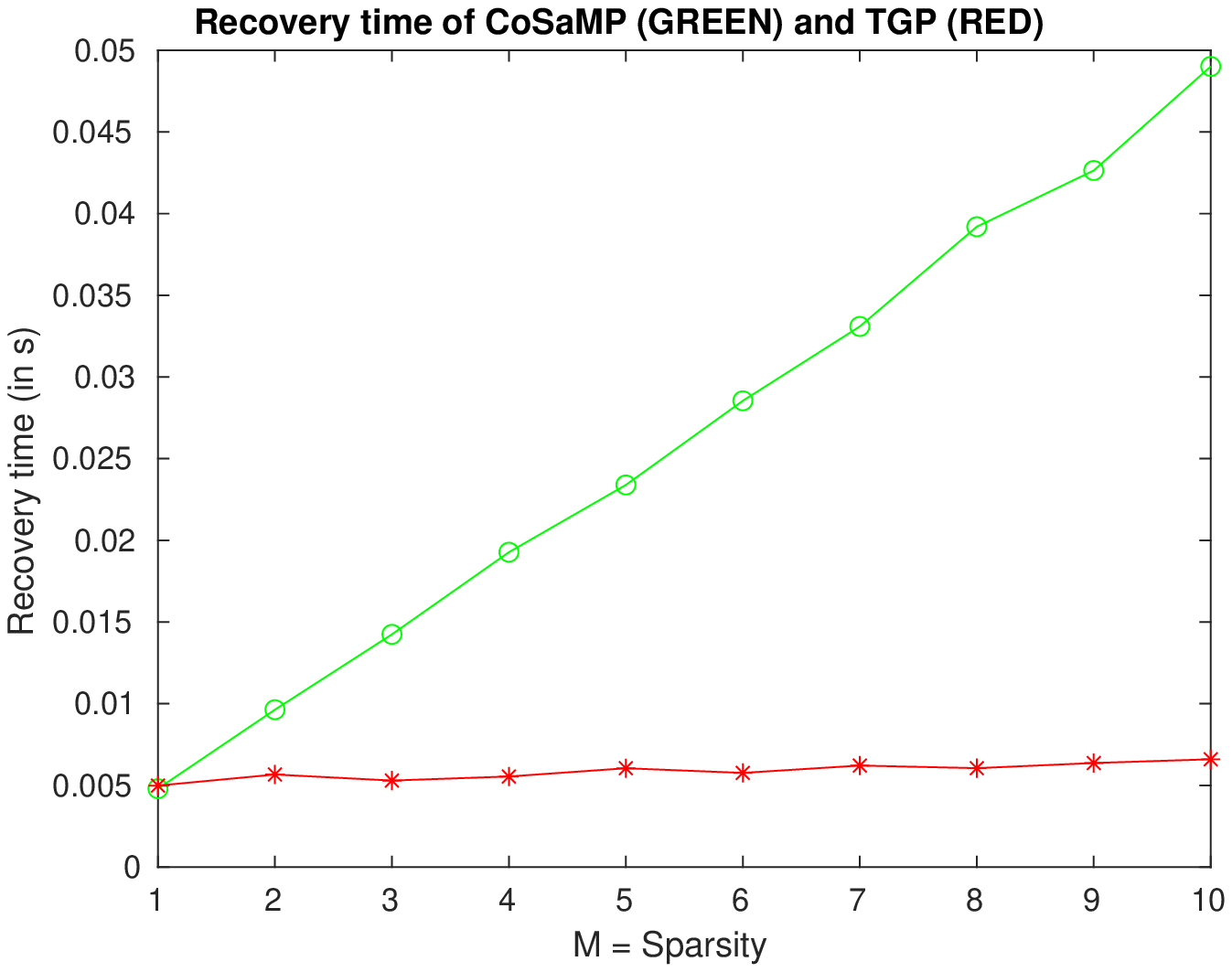}}  \subfloat{\includegraphics[width=0.333\textwidth]{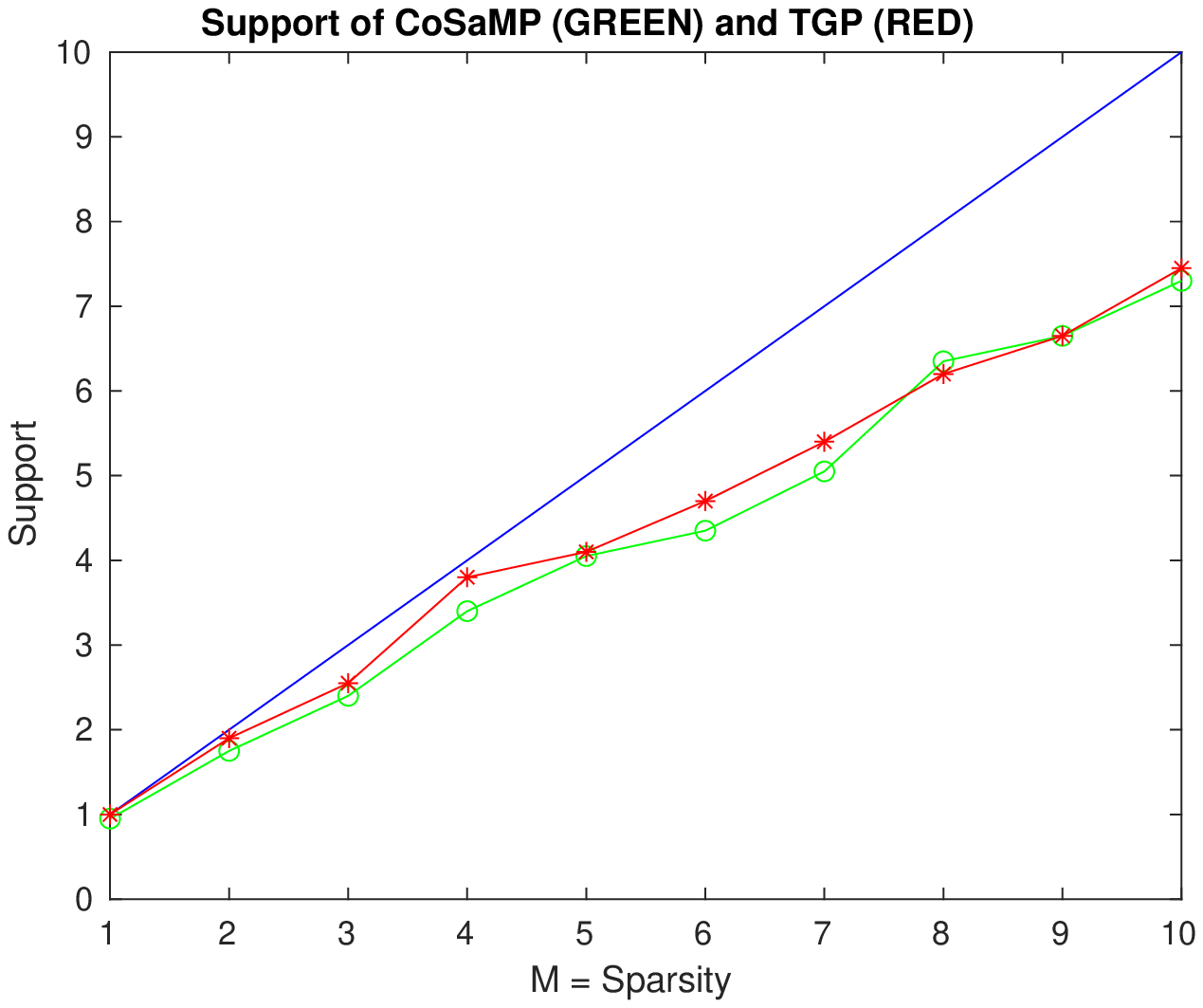}}
	\subfloat{\includegraphics[width=0.333\textwidth]{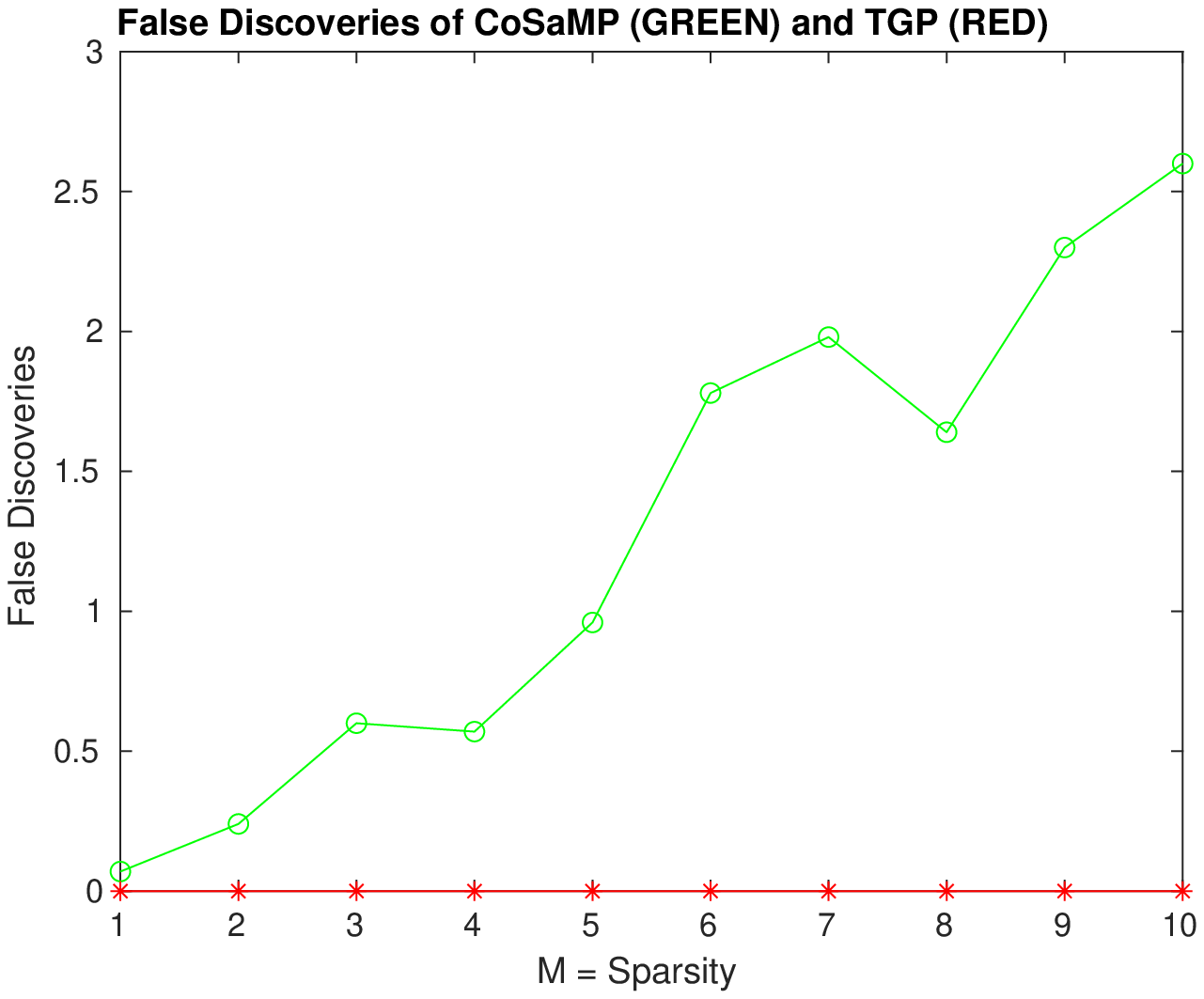}}
	\caption{$\ma$ is a Gaussian matrix, noise level is $\delta=1$. }
	\label{fig:gauss1}
\end{figure}

The settings for experiments are as follows. We generate a measurement matrix $\ma\in \mathbb R^{1600\times 3200}$ where each entry is a standard Gaussian random variable, an $M$-sparse vector $\vx\in\mathbb R^{1600}$ whose entries are randomly generated as $1+\chi$ where $\chi \sim N(0,1)$. We then input $\ma$ and $\bb = \ma\vx$ into both algorithms. We run both TGP and CoSaMP on the same data at different levels of noise $\delta = \|\ee\|_2/\|\bb\|_2$: $\delta=0$ (noiseless), $\delta=0.5$ (moderate level of noise), and $\delta=1$ (high level of noise). We vary $M$ from $1$ to $10$. For CoSaMP, at each $M$, we run the algorithm in $M$ iterations. For each $M$, we repeat the experiment for 20 times by regenerating $\vx$ and $\ee$. We want our sparse recovery algorithm to be fast, recover most of the support (or the whole support if the noise is not too large), and have zero false discoveries. Therefore, we 
measure the following parameters:
\begin{itemize}
	\item \textit{Recovery Time}: the time it takes the algorithm to complete, measured in seconds (less is better).
	\item \textit{Recovered Support:} the number of locations that are detected (more is better).
	\item \textit{False Discoveries}: the number of locations that are detected but not in the true support (less is better).
\end{itemize}
The resulting numbers are recorded and taken average over 20 times for each algorithm. The plots for these numbers are presented in Figures \ref{fig:gauss0}, \ref{fig:gauss05}, \ref{fig:gauss1} with red lines for TGP and green lines for CoSaMP. We can see the distinctions between the two algorithms. In every scenario, TGP runs faster than CoSaMP. In the noiseless case ($\delta=0$), TGP recovers the support exactly even with a high sparsity level. In the noisy case ($\delta>0$) and even at high sparsity levels, while CoSaMP starts having false discoveries, TGP does not and detects more of the support.

\newpage

\begin{figure}[ht] \centering 
	\captionsetup{width=.8\linewidth}
	\subfloat[$\ma$ is a Gaussian matrix.]{\includegraphics[width=0.45\textwidth]{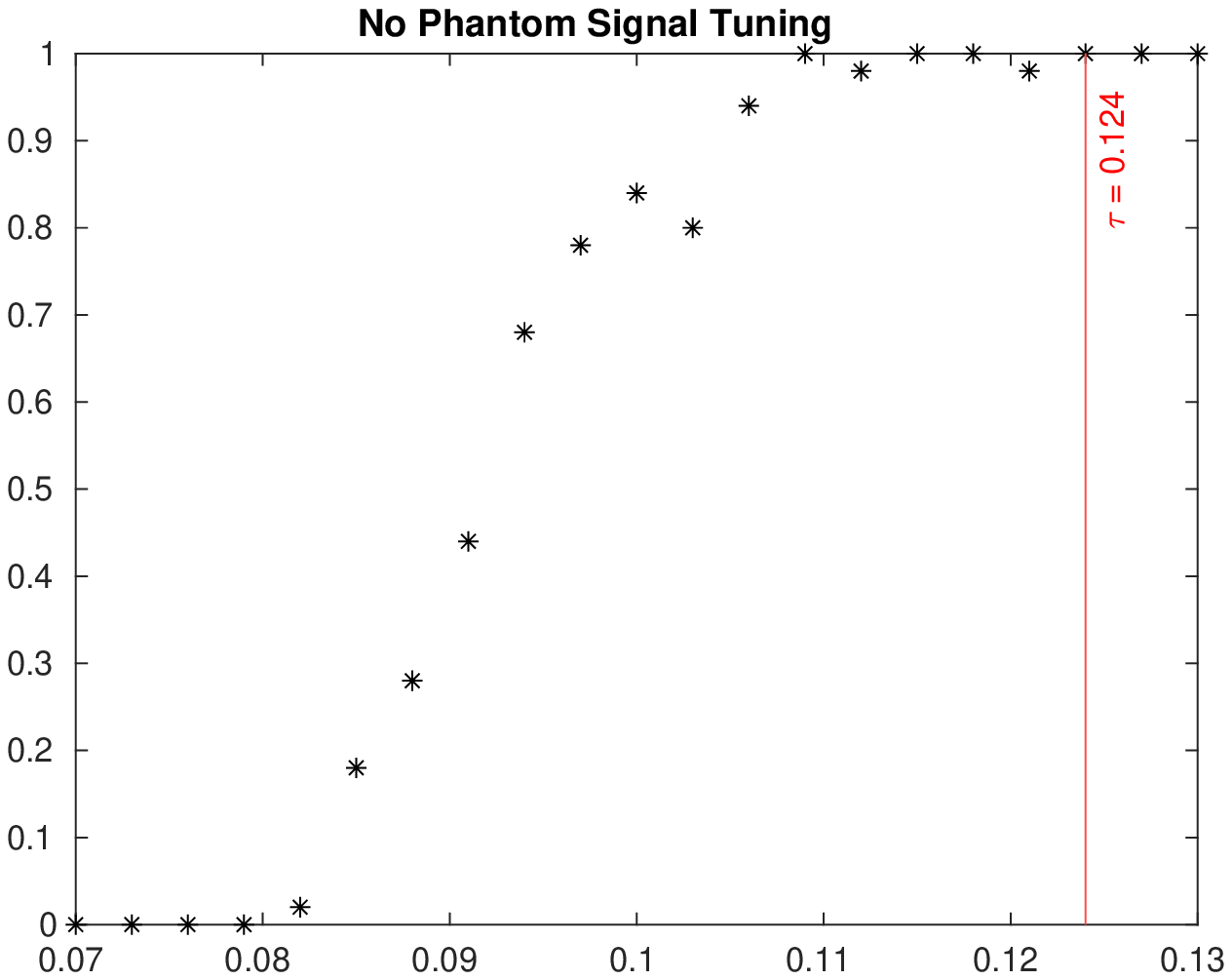}} 
	\subfloat[$\ma$ is a Partial Fourier matrix]{\includegraphics[width=0.45\textwidth]{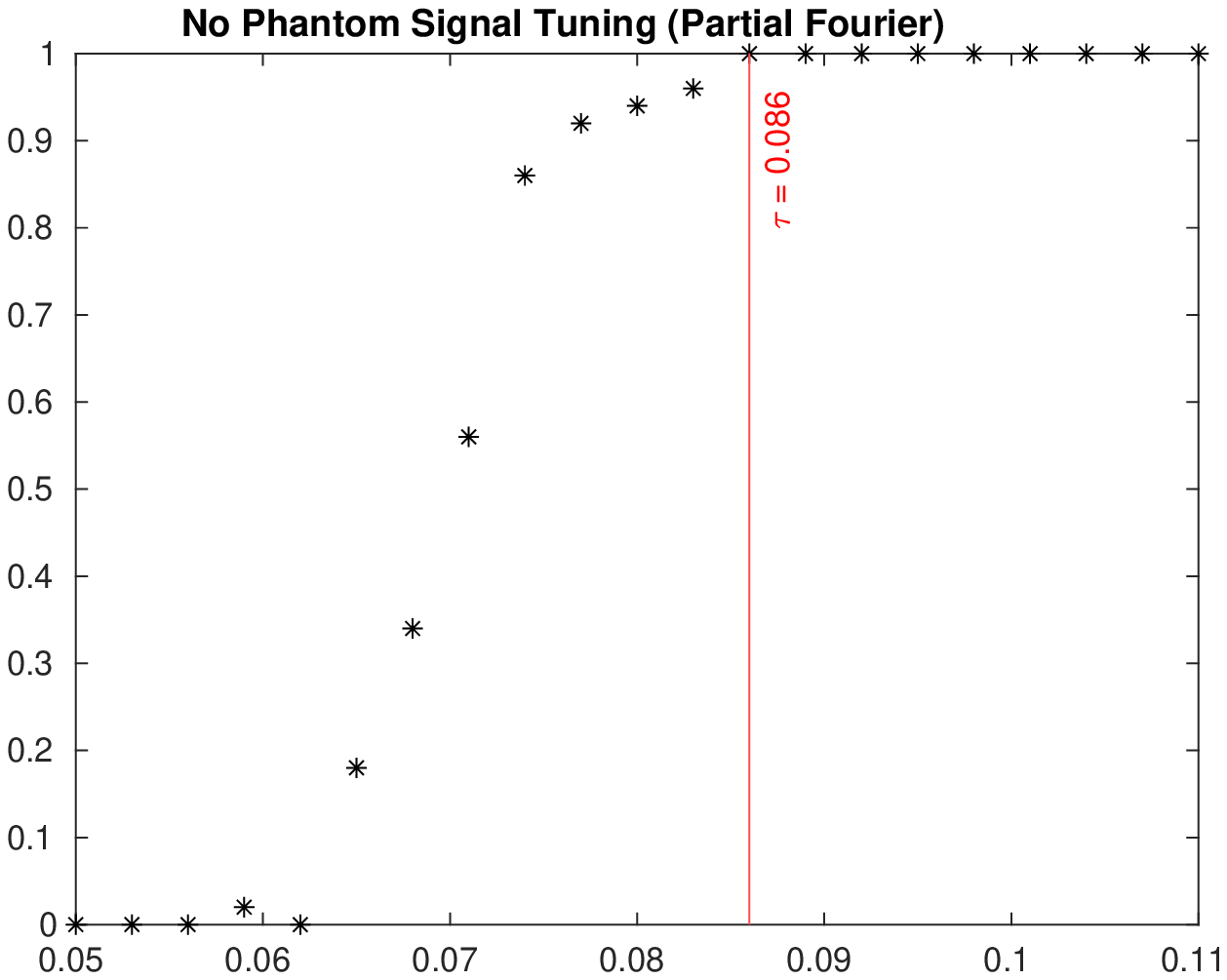}} 
	\caption{Transition diagrams of $\tau$ for No Phantom Signal test. Ordinate and abscissa are $\tau$ and rate of success.}
	\label{tautuning}
\end{figure}

We achieve optimal results for TGP, when we  calibrate $\tau$ numerically to be the smallest constant so that Theorem \ref{theorem:nophantom} holds. That is, we find numerically the smallest $\tau$, such that the algorithm outputs empty set 
when it is fed with pure noise measurements. We remark that choosing such $\tau$ does not require estimating the strength of noise $\|\ee\|_2$. We run TGP with input $\bb\in \mathbb R^{N}$ as a Gaussian vector, then vary $\tau$ from 0 to 1 with an increment of 0.003 until the algorithm only outputs empty set. For each $\tau$, we repeat the experiment by regenerating $\bb$ 50 times. We then record the rate of success ($\text{the number of successes}/50$). Each success is defined as each time the algorithm outputs empty support. We obtain the transition diagram for $\tau$ in Figure \ref{tautuning} (\textit{left}).

We see that starting from $\tau=0.124$, the rate of success remains 1, that is TGP outputs empty set 50 times out of 50 experiments. Therefore, we choose $\tau=0.124$ to be the thresholding parameter for TGP in the above experiments.

\begin{figure}[ht] \centering 
	\captionsetup{width=.8\linewidth}
	\subfloat{\includegraphics[width=0.333\textwidth]{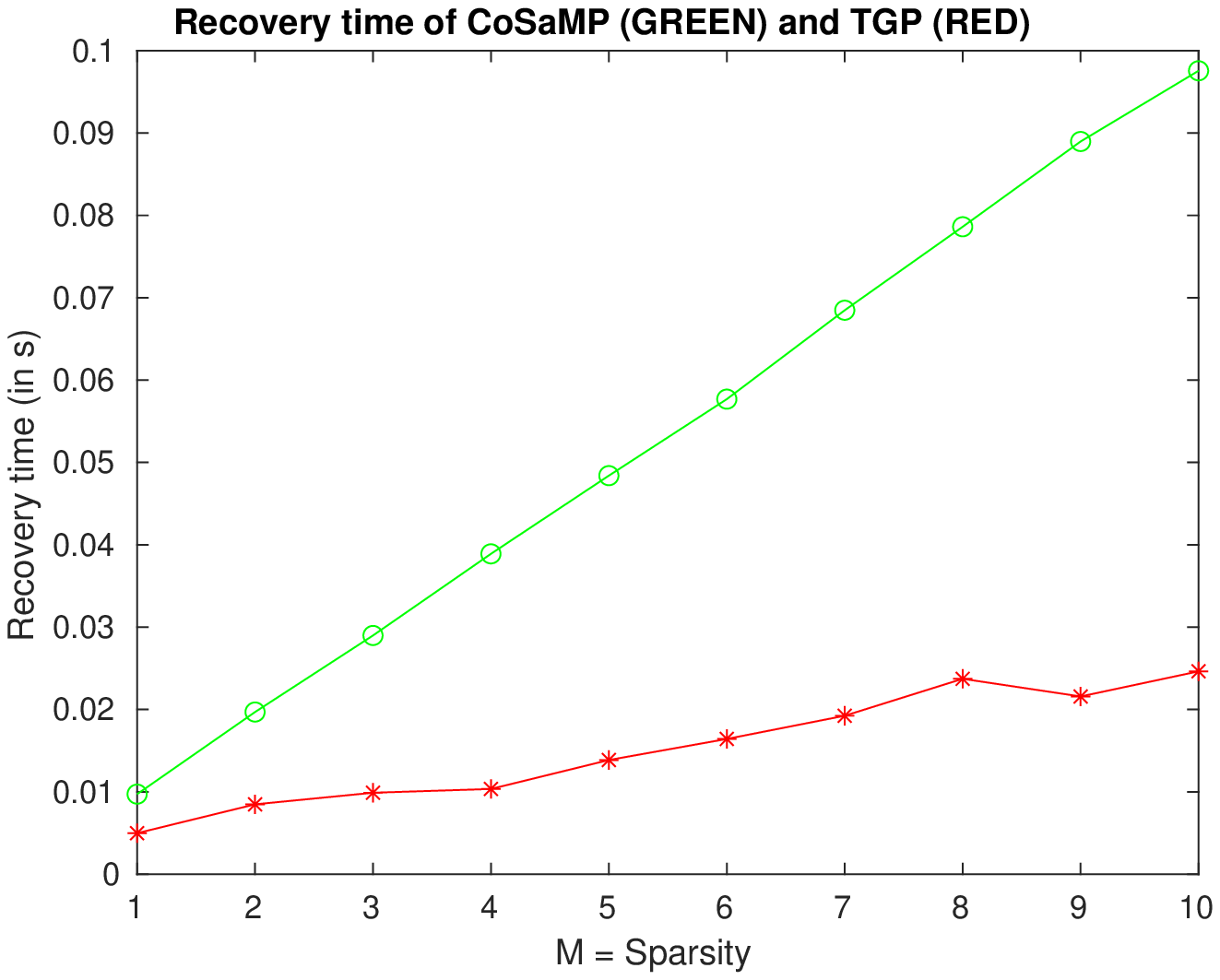}}  \subfloat{\includegraphics[width=0.333\textwidth]{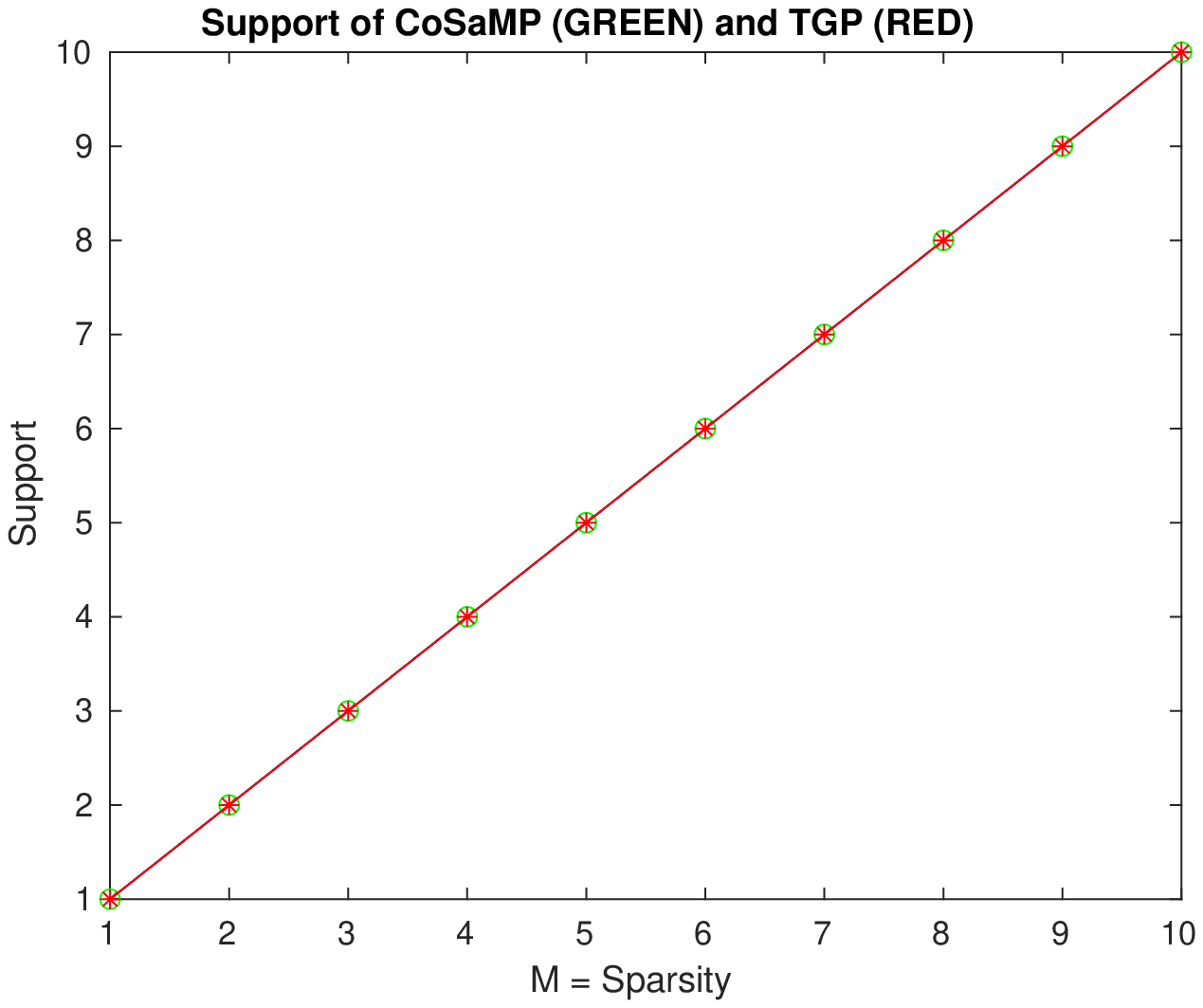}}
	\subfloat{\includegraphics[width=0.333\textwidth]{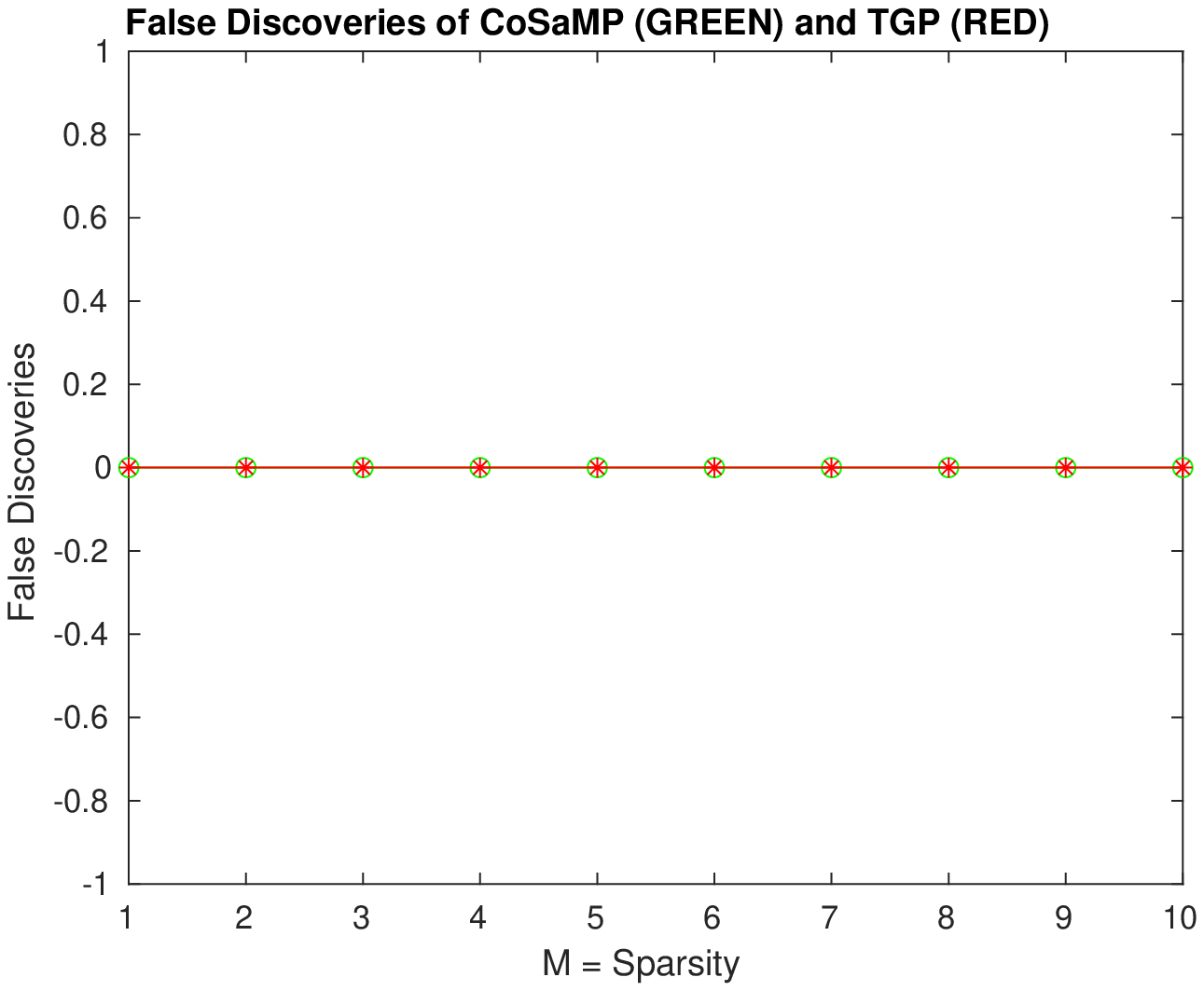}}
	\caption{$\ma$ is a Partial Fourier matrix, noise level is $\delta=0$. }
	\label{fig:fdt0}
\end{figure}

\begin{figure}[ht] \centering
	\captionsetup{width=.8\linewidth}
	\subfloat{\includegraphics[width=0.333\textwidth]{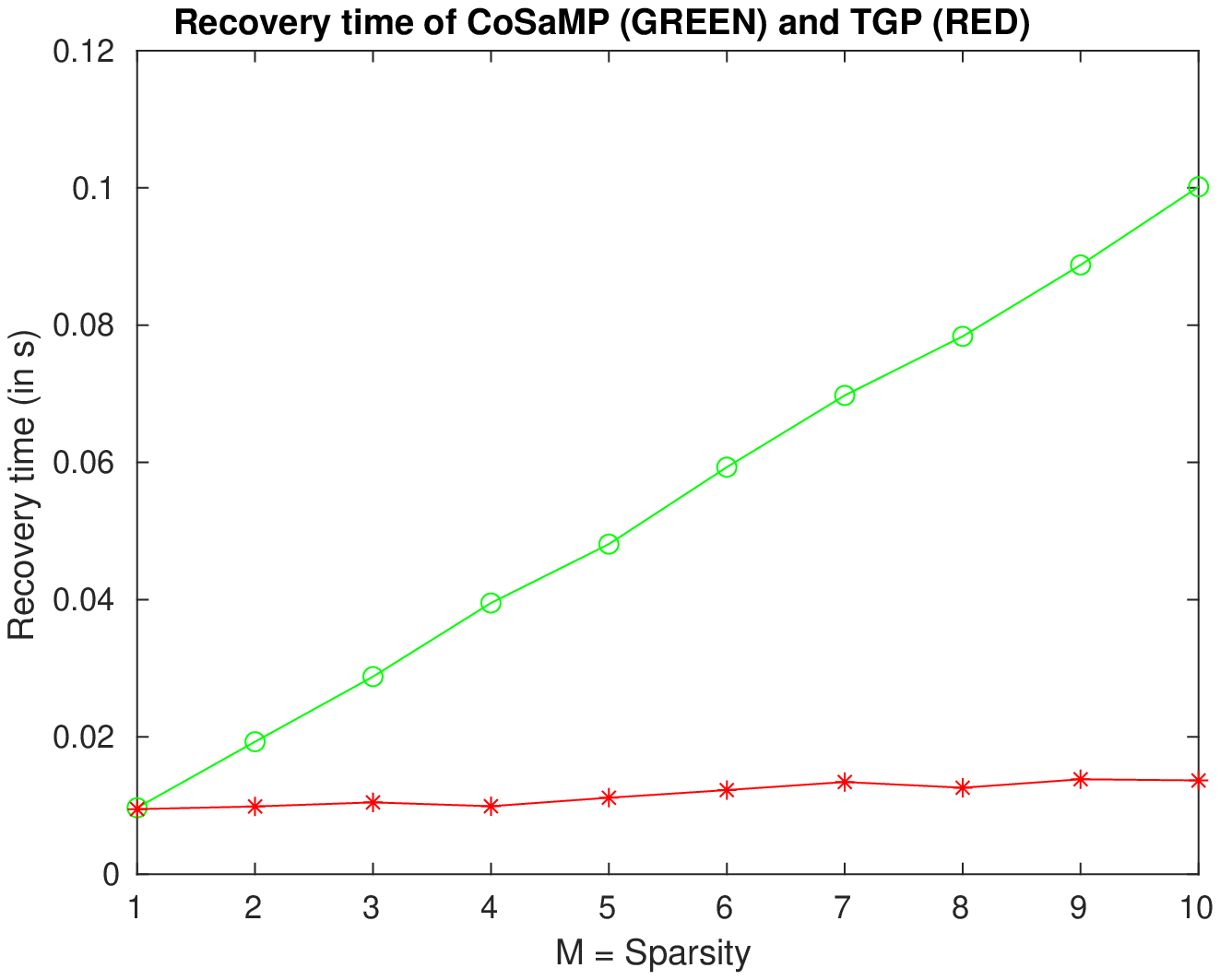}}  \subfloat{\includegraphics[width=0.333\textwidth]{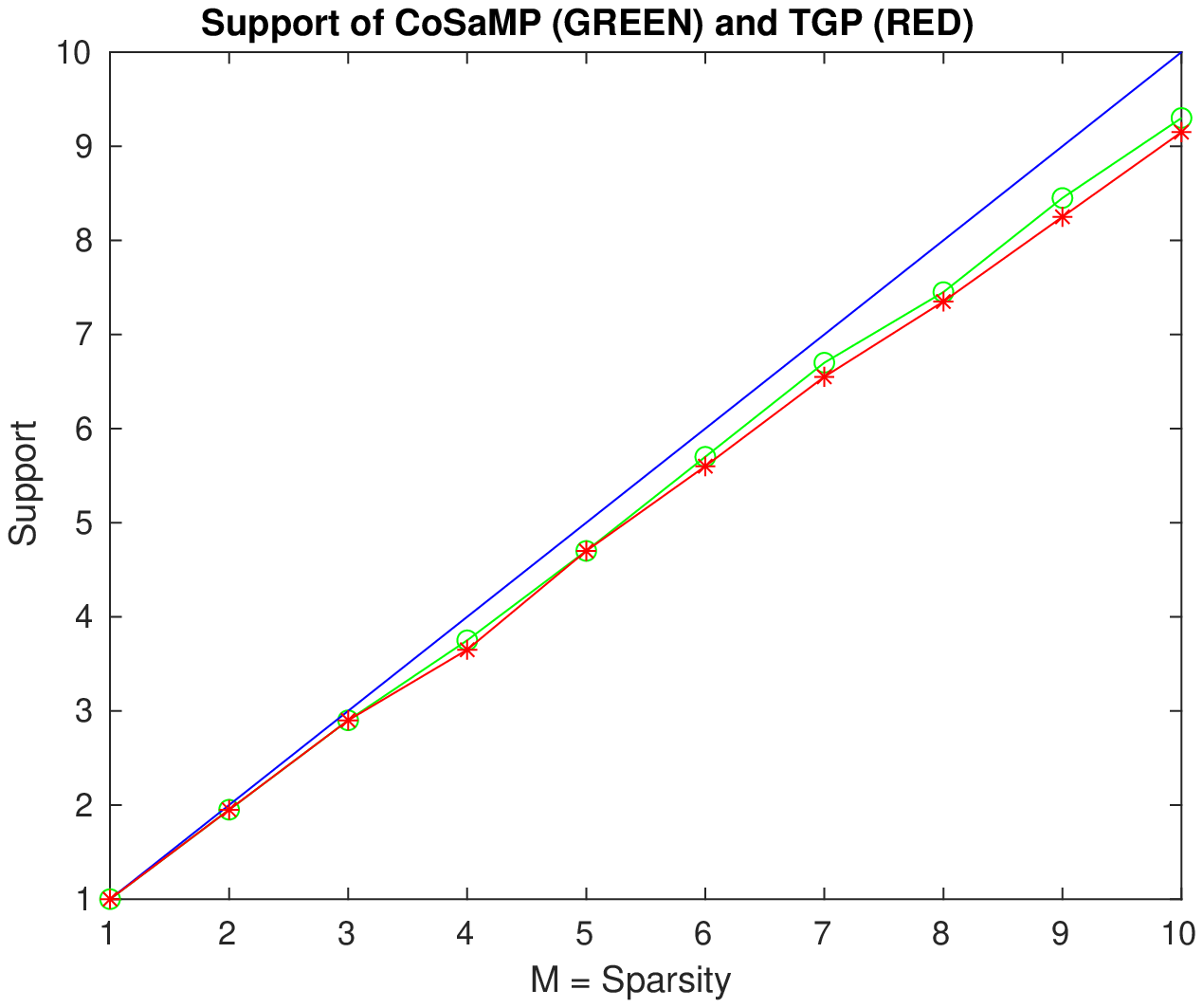}}
	\subfloat{\includegraphics[width=0.333\textwidth]{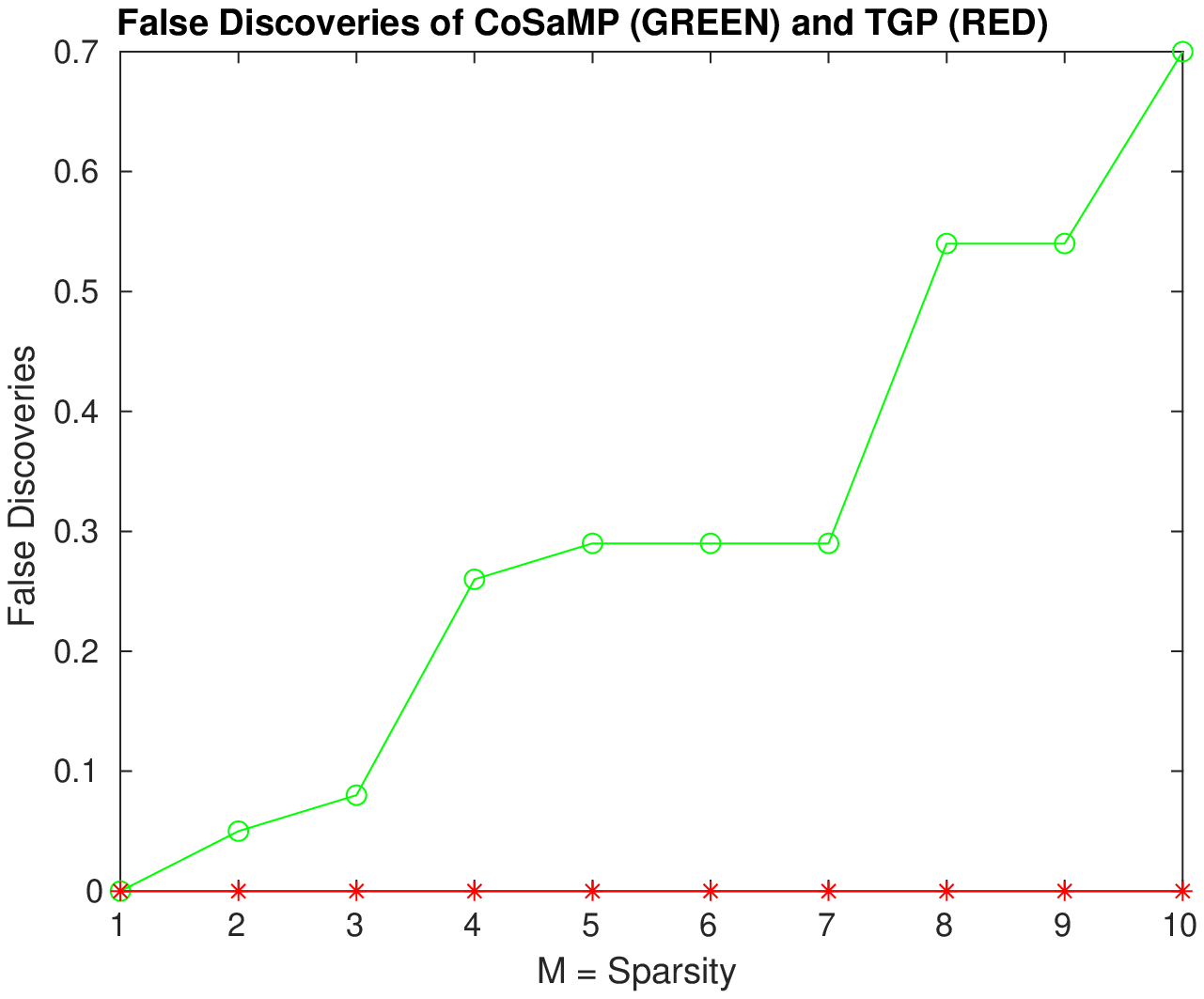}}
	\caption{$\ma$ is a Partial Fourier matrix, noise level is $\delta=0.5$.  }
	\label{fig:fdt05}
\end{figure}

\begin{figure}[ht] \centering 
	\captionsetup{width=.8\linewidth}
	\subfloat{\includegraphics[width=0.333\textwidth]{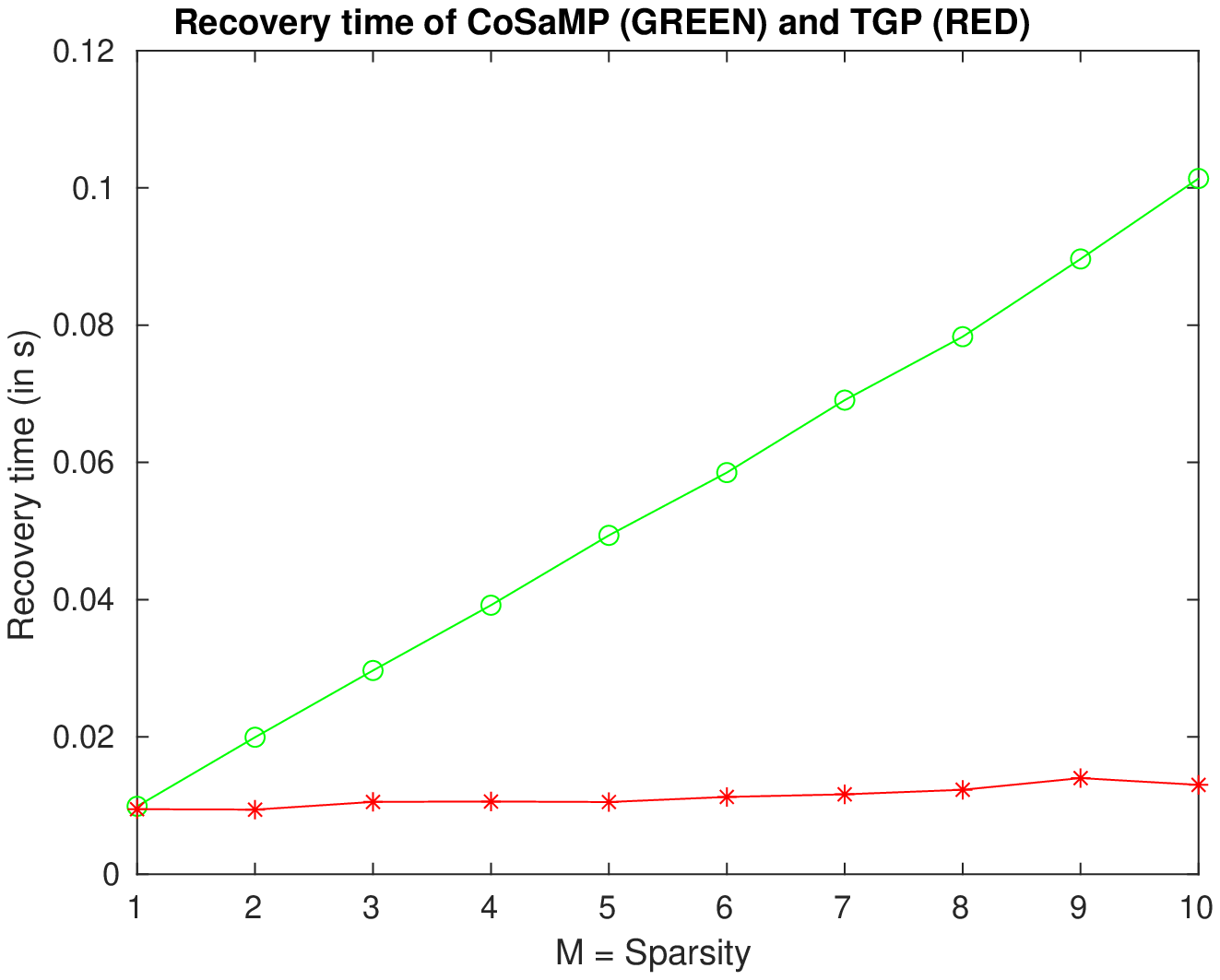}}  \subfloat{\includegraphics[width=0.333\textwidth]{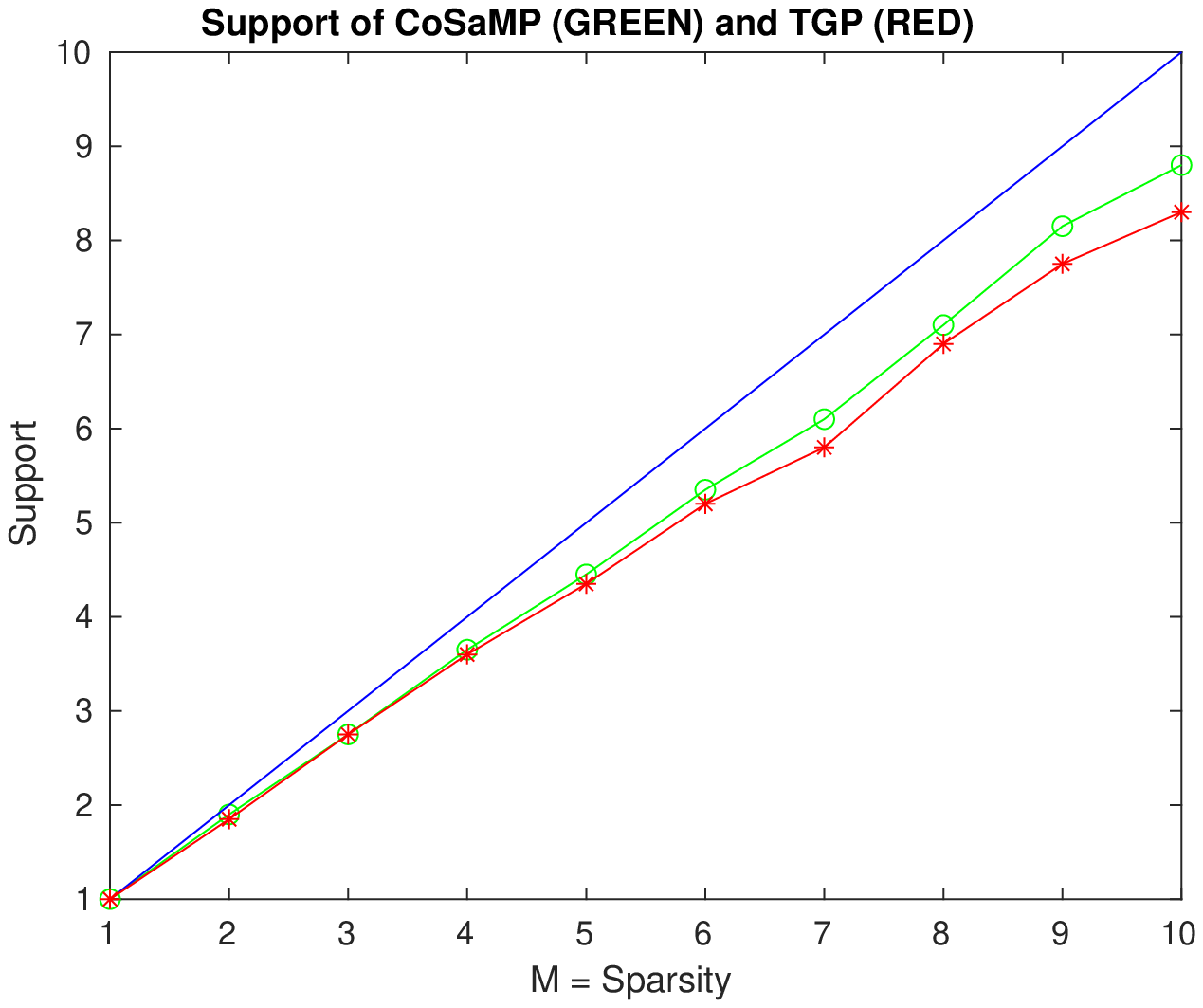}}
	\subfloat{\includegraphics[width=0.333\textwidth]{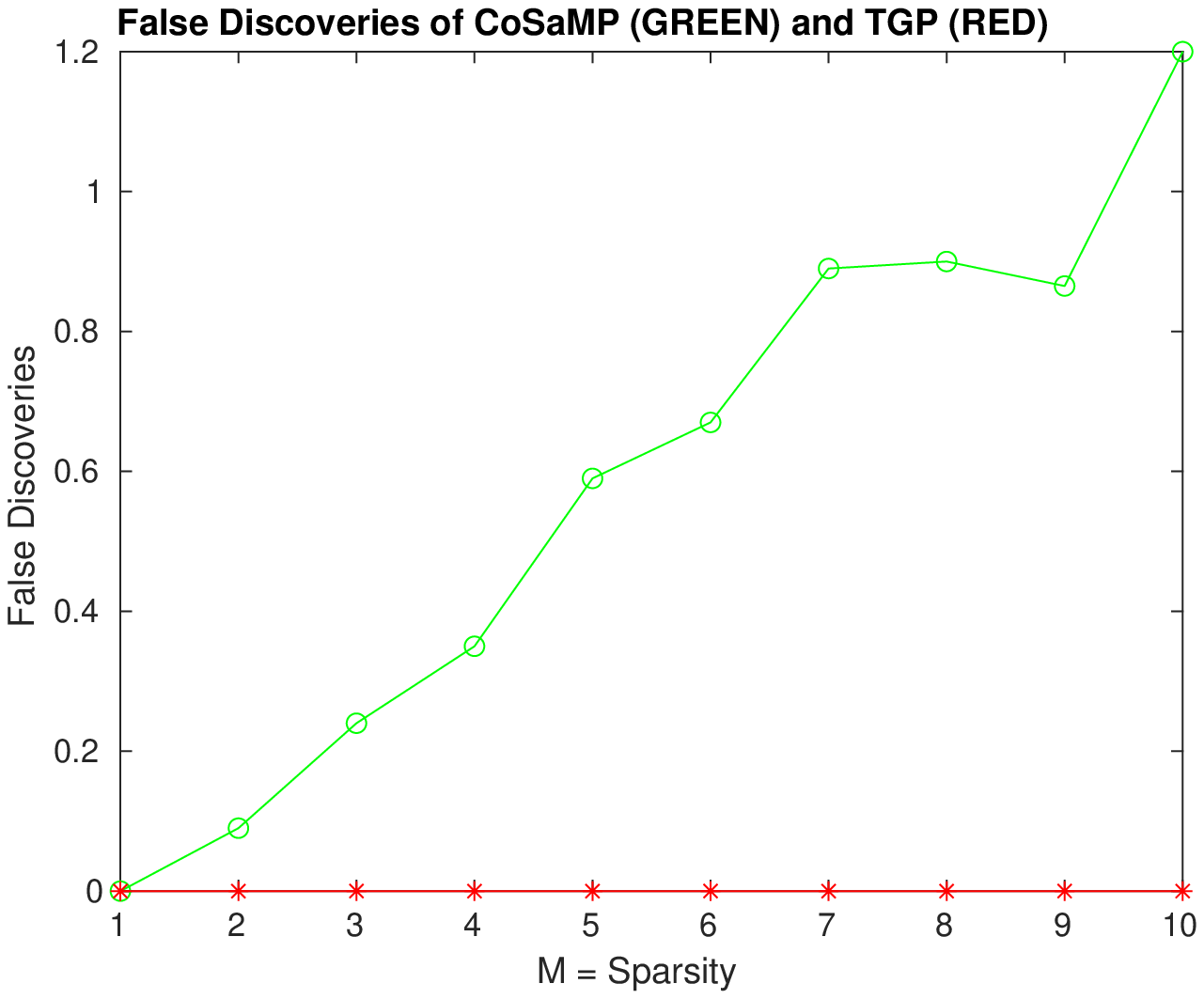}}
	\caption{$\ma$ is a Partial Fourier matrix, noise level is $\delta=1$.  }
	\label{fig:fdt1}
\end{figure}

We also run the same experiments for the case of Partial Fourier matrix where $\ma$ is a uniformly random set of $N=1600$ rows drawn from the $K \times K$ ($K=3200$) unitary discrete Fourier transform (DFT). In these experiments, we choose $\tau=0.086$ by calibrating the parameter using the same procedure as in the Gaussian case. The transition diagram for $\tau$, in this case, is in Figure \ref{tautuning} (\textit{right}). The performance plots are presented in Figures \ref{fig:fdt0}, \ref{fig:fdt05}, \ref{fig:fdt1}. TGP recovers slightly less of the support than CoSaMP but has no false discoveries even at high levels of sparsity and noise.

\newpage

\section{Proofs of Main Results}\label{section3}
In this section, we prove the main theorems introduced in Section \ref{sec:maintheorem}. We then end with an discussion on related questions and future directions. 
\subsection{No Phantom Signal}
The proof of Theorem \ref{theorem:nophantom} is as follows.
\begin{proof}
	If we show that
	\begin{equation}\label{thm1_1}
	\max\limits_{1\le i\le K}|\la \va_i,\ee\ra|\le \tau \|\ee\|_2
	\end{equation}
p	with probability $1-2/N^{\kappa}$, then after the \textbf{Thresholding} step, $\vx^{1}$ is always a zero vector. Without loss of generality, we assume $\|\ee\|_2=1$. By independence, we have that $\PP\left(|\la \va_i,\ee\ra|\ge t/\sqrt{N} \right)\le 2\exp(-t^2/2)$ for each $\va_i$. Here, we make use of the fact that uniformly distributed vectors in high dimension behave like Gaussian \cite{Vershynin2018}. Consequently, the union bound gives
	\[
	\PP\left(\max\limits_{1\le i\le K}|\la \va_i,\ee\ra|\ge t/\sqrt{N}\right)\le 2K\exp(-t^2/2)\le 2N^{\gamma}\exp(-t^2/2).
	\]
	For $t=c_0\sqrt{\log N}$, the right-hand side becomes $2N^{\gamma-c_0^2/2}$. Therefore,
	\begin{equation}
	\PP\left( \max\limits_{1\le i\le K}|\la \va_i,\ee\ra|\le c_0\frac{\sqrt{\log N}}{\sqrt{N}}\right)\ge 1-\frac{2}{N^{c_0^2/2-\gamma}}\label{eq:rotational}.
	\end{equation}
	Choosing $c_0=\sqrt{2(\gamma+\kappa)}$, we deduce that  inequality~\eqref{thm1_1} holds for $\tau\ge c_0\sqrt{\log N}/\sqrt{N}$.  
\end{proof}

\subsection{No False Discoveries}
The proof of Theorem \ref{theorem:nofalse} is as follows.
\begin{proof} Consider the event
	\[
	\mathcal{O} = \left\{ \max\limits_{1\le i\le K}|\la \va_i,\ee\ra|\le c_0\frac{{\sqrt{\log N}}}{\sqrt{N}} \|\ee\|_2 \right\}.
	\]
	According to \eqref{eq:rotational}, $\mathcal{O}$ holds with probability $1-2/N^{\kappa}$. Suppose the event $\mathcal{O}$ occurs, and the following analysis is deterministic on this event.
	
	Without loss of generality, let us assume that $\bb = \sum_{i=1}^Mx_i\va_i+\ee$. Our proof is by induction on the number of iterations. Consider the first iteration.  We want to show that 
	\[
	|\la \va_j, \bb\ra| \le \tau \|\bb\|_2, \mbox{ for all } j \not\in \supp(\vx).
	\]
	Pick $j \not\in \supp(\vx)$, and let $P = [\va_1,\ldots,\va_M,\ee/\|\ee\|_2]$, that is a matrix whose columns are $\va_1$, $\ldots$, $\va_M$, $\ee/\|\ee\|_2$. It suffices to show that the length of the orthogonal projection of  $\va_j$
	onto the vector space spanned by columns of $P$ never exceeds $\tau$. Indeed, if that is the case, then by Cauchy-Schwarz,
	\[
	|\la \va_j,\bb\ra| = |\la P(P^*P)^{-1}P^*\va_j,\bb\ra|\le\|P(P^*P)^{-1}P^*\va_j\|_2\|\bb\|_2\le \tau \|\bb\|_2.
	\]
	Thus it suffices to show
	\begin{equation}\label{eq:form1}
	\|P(P^*P)^{-1}P^*\va_j\|_2^2 \le \tau^2  \mbox{ for all } j \not\in \supp(\vx).
	\end{equation}
	We have that 
	\[
	\|P(P^*P)^{-1}P^*\va_j\|_2^2 = \la P^*\va_j,(P^*P)^{-1}P^*\va_j \ra \le \|(P^*P)^{-1}\|\|P^*\va_j\|_2^2.
	\]
	Using condition $M\le 1/(4\mu)$ and $|\la \va_j,\ee/\|\ee\|_2\ra|\le c_0\sqrt{\log N}/\sqrt{N}$,  we have 
	\[
	\|P^*\va_j\|_2^2 = \sum\limits_{i=1}^M|\la \va_j,\va_i\ra|^2 + |\la\va_j,\ee/\|\ee\|_2\ra|^2\le M\cdot \mu^2+\frac{c_0^2\log N}{N} \le \frac{\mu}{4}+\frac{c_0^2\log N}{N}.
	\]
	
	To estimate the operator norm $\|(P^*P)^{-1}\|$ we note that $P^*P-I$ is an $(M+1)\times(M+1)$ matrix whose diagonal entries $a_{ii}$ are zero and off-diagonal entries $a_{ik}$ ($i\neq k$) are no bigger than $1/(4M)$. By Gershgorin circle theorem 
	(see, for example,~\cite{Golub2012}), we obtain
	\[
	\|P^*P-I\| \le M\cdot \frac{1}{4M}=\frac{1}{4}.
	\]
	This implies that $\|P^*P\|\ge 1-1/4=3/4$. Hence, $\|(P^*P)^{-1}\|\le 4/3$. Thereofore we obtain
	\begin{equation}\label{eq:form}
	\|P(P^*P)^{-1}P^*\va_j\|_2^2 \le \frac{4}{3}\cdot \left( \frac{\mu}{4}+\frac{c_0^2\log N}{N} \right)=\tau^2  \mbox{ for all } j \not\in \supp(\vx).
	\end{equation}
	
	We proceed to the induction step. Suppose  we have already recovered $\Omega^n \subset \supp(\vx)$ during the previous $n$  iterations and did not have any false discoveries. We now show that in the next iteration the algorithm will not make any false discoveries.
	Denote, for brevity, $\Omega=\Omega^n$, and $\bb=\bb^n$. Suppose that $|\Omega|=k$ where $0\le k\le M$. It suffices to show that
	\[
	|\la \va_j,(I-\ma_{\Omega}(\ma_{\Omega}^*\ma_{\Omega})^{-1}\ma_{\Omega}^*)\bb\ra|\le \tau \|(I-\ma_{\Omega}(\ma_{\Omega}^*\ma_{\Omega})^{-1}\ma_{\Omega}^*)\bb\|_2  \mbox{ for all } j \not\in \supp(\vx).
	\]
	We have the following orthogonal decomposition
	\[
	\va_j = (I- P(P^*P)^{-1}P^*)\va_j+P(P^*P)^{-1}P^*\va_j \mbox{ for any } j.
	\]
	Pick $j \not\in \supp(\vx)$, and observe that $(I - P(P^*P)^{-1}P^*)\va_j$ is orthogonal to $\bb$. Moreover, since the image of the orthogonal projection matrix $(I-\ma_{\Omega}(\ma_{\Omega}^*\ma_{\Omega})^{-1}\ma_{\Omega}^*)$ is the vector space that is orthogonal to the space spanned by columns of $\ma_{\Omega}$, and columns of $\ma_{\Omega}$ are also columns of $P$, we must have that 
	\[ 
	(I - P(P^*P)^{-1}P^*)\va_j \perp (I-\ma_{\Omega}(\ma_{\Omega}^*\ma_{\Omega})^{-1}\ma_{\Omega}^*)\bb. 
	\]
	From this, we obtain
	\[
	\la \va_j, (I-\ma_{\Omega}(\ma_{\Omega}^*\ma_{\Omega})^{-1}\ma_{\Omega}^*)\bb\ra = \la P(P^*P)^{-1}P^*\va_j, (I-\ma_{\Omega}(\ma_{\Omega}^*\ma_{\Omega})^{-1}\ma_{\Omega}^*)\bb\ra.
	\]
	Then, by Cauchy-Schwarz inequality, we have
	\begin{align*}
	| \la P(P^*P)^{-1}P^*\va_j,(I-\ma_{\Omega}(\ma_{\Omega}^*\ma_{\Omega})^{-1}\ma_{\Omega}^*)\bb\ra|&\le \|P(P^*P)^{-1}P^*\va_j\|_2 \|(I-\ma_{\Omega}(\ma_{\Omega}^*\ma_{\Omega})^{-1}\ma_{\Omega}^*)\bb\|_2\\&\le \tau \|(I-\ma_{\Omega}(\ma_{\Omega}^*\ma_{\Omega})^{-1}\ma_{\Omega}^*)\bb\|_2.
	\end{align*}
	The proof is complete.
\end{proof}
\subsection{Exact Recovery }
The proof of Theorem \ref{theorem:exact} is as follows.
\begin{proof} Similarly to Theorem \ref{theorem:nofalse}, we consider the event $\mathcal{O}$ holds.
	\bigskip
	
	The objective is to demonstrate that at least one nonzero entry of $\vx$ is detected at every iteration. It suffices to look at the nonzero entry of $\vx$ with the largest magnitude. 
	Consider the $n$-th iteration, and assume $\Omega:=\Omega^n$ is strictly contained in the support of $\vx$. According to Theorem \ref{theorem:nofalse}, $|\Omega|\le M$. Let $\Omega^{c}=\supp(\vx)\backslash \Omega$, the set of undetected indices. 
	Without loss of generality, assume that the first entry $x_1$ has is the nonzero entry with the largest magnitude among $\{x_k, k\in\Omega^c\}$. We want to show that 
	\begin{equation}\label{supportineq}
	|\la \va_1,\bb^{n}\ra|>\tau \|\bb^{n}\|_2,
	\end{equation}
	so then the index $1$ will be included in $\Omega^{n+1}$.
	
	Decompose $\bb$ into $\va_1x_1 + \ma_{\Omega^{c}\backslash\{1\}}\vx_{\Omega^c\backslash\{1\}}+\ma_{\Omega}\vx_{\Omega}+\ee.$ Notice that, by projecting $\bb$ onto the orthogonal complement of the vector space spanned by $\{\va_k:k\in \Omega\}$, we have  
	\[
	\bb^{n}=\bb-\ma_{\Omega}\ma_{\Omega}^{\dag}\bb = \va_1 x_1+\ma_{\Omega^c\backslash\{1\}}\vx_{\Omega^c\backslash\{1\}} -\ma_{\Omega}\ma_{\Omega}^{\dag}(\ma_{\Omega^c}\vx_{\Omega^c})+(\ee-\ma_{\Omega}\ma_{\Omega}^{\dag}\ee).
	\]
	For convenience, let us now set $\vv:=\ma_{\Omega^c\backslash\{1\}}\vx_{\Omega^c\backslash\{1\}} -\ma_{\Omega}\ma_{\Omega}^{\dag}(\ma_{\Omega^c}\vx_{\Omega^c})+(\ee-\ma_{\Omega}\ma_{\Omega}^{\dag}\ee)$. We then observe that 
	\begin{align}
	|\la \va_1,\bb^n\ra|-\tau \|\bb^n\|_2&= 	|\la \va_1,\va_1x_1+\vv\ra|^2- \tau^2\|\va_1x_1+\vv\|_2^2\nonumber\\
	&=	(1-\tau^2)\left|x_1+\la \va_1,\vv\ra\right|^2+\tau^2|\la \va_1,\vv\ra|^2- \tau^2\|\vv\|_2^2 \nonumber\\
	&> (1-\tau^2)\left|x_1+\la \va_1,\vv\ra\right|^2- \tau^2\|\vv\|_2^2.\nonumber
	\end{align}
	By triangle inequality, we have $|x_1+\la \va_1,\vv\ra|\ge |x_1|-|\la\va_1,\vv\ra|$. Therefore, it suffices to show that
	\begin{equation}\label{eq:noiseless}
	(1-\tau^2)\left(|x_1|-|\la \va_1,\vv\ra|\right)^2>\tau^2\|\vv\|_2^2.
	\end{equation}
	Let us estimate $|\la \va_1,\vv\ra|$. We have that
	\[ 
	\begin{aligned}
	|\la \va_1,\vv\ra|\le |\la \va_1,\ma_{\Omega^c\backslash\{1\}}\vx_{\Omega^c\backslash\{1\}}\ra|+|\la\va_1, \ma_{\Omega}\ma_{\Omega}^{\dag}(\ma_{\Omega^c}\vx_{\Omega^c})\ra| + |\la \va_1,\ee-\ma_{\Omega}\ma_{\Omega}^{\dag}\ee\ra|. 
	\end{aligned}
	\]
	We estimate each term on the right-hand side as follows. 
	\begin{itemize}
		\item For $|\la \va_1,\ma_{\Omega^c\backslash\{1\}}\vx_{\Omega^c\backslash\{1\}}\ra|$:
		
		By using condition \eqref{mutualcondition} that $\max\limits_{i\neq j}|\la \va_i,\va_j\ra|\le 1/(4M)$ and knowing $|x_1|$ being the largest among $\{|x_k|,k\in \Omega ^c\}$, we have that 
		\[|\la \va_1,\ma_{\Omega^c\backslash\{1\}}\vx_{\Omega^c\backslash\{1\}}\ra|\le \frac{(M-1)|x_1|}{4M} <\frac{|x_1|}{4}.\]
		\item For $|\la\va_1, \ma_{\Omega}\ma_{\Omega}^{\dag}(\ma_{\Omega^c}\vx_{\Omega^c})\ra| $:
		
		We have that 
		\[ 
		\begin{aligned} 
		|\la \va_1,\ma_{\Omega}\ma_{\Omega}^{\dag}(\ma_{\Omega^c}\vx_{\Omega^c})\ra|&=|\la \ma_{\Omega}^*\va_1,(\ma_{\Omega}^*\ma_{\Omega})^{-1}\ma_{\Omega}^*\ma_{\Omega^c}\vx_{\Omega^c}\ra|\\
		&\le \|\ma_{\Omega}^*\va_1\|_2\|(\ma_{\Omega}^*\ma_{\Omega})^{-1}\|\|\ma^*_{\Omega}\ma_{\Omega^c}\vx_{\Omega^c}\|_2.\\
		\end{aligned}
		\]
		We will estimate each term in the product. The first term is as follows 
		\[
		\|\ma_{\Omega}^*\va_1\|_2=\sqrt{\sum\limits_{k\in \Omega}|\la \va_k,\va_1\ra|^2}\le \sqrt{M\cdot\frac{1}{16M^2}}=\frac{1}{4\sqrt{M}}.
		\]
		The second term is estimated by using Gershgorin circle theorem, which yields $\|(\ma_{\Omega}^*\ma_{\Omega})^{-1}\|\le 4/3$. For the third term, by using incoherence again, each entry of the vector $\ma_{\Omega}^*\ma_{\Omega^c}\vx_{\Omega^c}$ is less than, in absolute value,
		\[
		\frac{1}{4M}(|x_1|+\ldots+|x_{|\Omega^c|}|)\le \frac{1}{4M}\cdot M|x_1|=\frac{|x_1|}{4}.
		\]
		Therefore, we have that 
		\begin{equation}
		\|\ma_{\Omega}^*\ma_{\Omega^c}\vx_{\Omega^c}\|_2\le \frac{\sqrt{M}|x_1|}{4}.\label{eq:estimate}
		\end{equation} Overall, we obtain the following bound
		\[
		|\la\va_1,\ma_{\Omega}\ma_{\Omega}^{\dag}(\ma_{\Omega^c}\vx_{\Omega^c})\ra|\le \frac{1}{4\sqrt{M}}\cdot \frac{4}{3}\cdot \frac{\sqrt{M}|x_1|}{4}=\frac{|x_1|}{12}.
		\]
		\item For $|\la \va_1,\ee-\ma_{\Omega}\ma_{\Omega}^{\dag}\ee\ra|$:
		
		We will show that 
		\[
		|\la \va_1,\ee-\ma_{\Omega}\ma_{\Omega}^{\dag}\ee\ra|\le \frac{4}{3}\tau\|\ee\|_2.
		\]
		If $|\Omega| = 0$, it is true since the event $\mathcal O$ holds. We then consider the case when $1\le |\Omega|\le M$. Suppose that $\ma_{\Omega}\ma_{\Omega}^{\dag}\va_1 = \sum_{j\in \Omega}\xi_j \va_j$. Let $j^*\in \Omega$ be the index such that $|\xi_{j^*}| = \max_{j\in \Omega}|\xi_j|=\| \xxi\|_{\infty}$. By using $\max\limits_{i\neq j}|\la \va_i,\va_j\ra|\le 1/(4M)$, we have that 
		\[
		\displaystyle\frac{1}{4M}\ge \left |\left\langle \va_1,\va_{j^*}\right\rangle\right |= |\la \ma_{\Omega}\ma_{\Omega}^{\dag}\va_1,\va_{j^*}\ra|= |\la  \sum_{j\in \Omega}\xi_j \va_j,\va_{j^*}\ra|\ge \| \xxi\|_{\infty}\left(1-\frac{|\Omega|-1}{4M} \right)\ge \frac{3}{4}\| \xxi\|_{\infty}.
		\]
		Therefore, $\| \xxi\|_{\infty}\le 1/(3M)$, which implies that $\| \xxi\|_{1}\le M\cdot \|\xxi\|_{\infty}\le 1/3$. Consequently, we have
		\begin{align*}
		|\la \va_1,\ee-\ma_{\Omega}\ma_{\Omega}^{\dag}\ee\ra|&= |\la (I-\ma_{\Omega}\ma_{\Omega}^{\dag})\va_1,\ee\ra| \\
		&\le |\la\va_1,\ee\ra|+ \sum_{j\in\Omega}|\xi_j||\la \va_j,\ee\ra|\\
		&\le \tau \|\ee\|_2+\|\xxi\|_1\cdot \tau \|\ee\|_2\\
		&\le \frac{4}{3}\tau\|\ee\|_2.
		\end{align*}
	\end{itemize}

	Combining the above estimates, we obtain
	\[
	|x_1|-|\la \va_1,\vv\ra|\ge |x_1|-\frac{|x_1|}{4}-\frac{|x_1|}{12}-\frac{4}{3}\tau\|\ee\|_2=\frac{2|x_1|}{3}-\frac{4}{3}\tau\|\ee\|_2.
	\] 
	Therefore, the left-hand side of \eqref{eq:noiseless} is bigger than $(1-\tau^2)(2|x_1|/3-4\tau\|\ee\|_2/3)^2$. Now, we look at the right-hand side. We have that 
	\[ 
	\|\vv\|_2\le \|\ma_{\Omega^c\backslash\{1\}}\vx_{\Omega^c\backslash\{1\}}\|_2+\|\ma_{\Omega}\ma_{\Omega}^{\dag}(\ma_{\Omega^c}\vx_{\Omega^c})\|_2+\|\ee-\ma_{\Omega}\ma_{\Omega}^{\dag}\ee\|_2.
	\]
	We estimate each term as follows
	\begin{itemize}
		\item For $\|\ma_{\Omega^c\backslash\{1\}}\vx_{\Omega^c\backslash\{1\}}\|_2$:
		
		We have that 
		\begin{align*}
		\|\ma_{\Omega^c\backslash\{1\}}\vx_{\Omega^c\backslash\{1\}}\|_2^2& = \sum\limits_{i\in \Omega^c\backslash\{1\}} |x_i|^2+2\sum\limits_{i<j,i,j\in \Omega^c\backslash\{1\}}\text{Re} (\la \va_i,\va_j\ra \bar{x}_ix_j)\\
		&\le (M-1)|x_1|^2+2\cdot \frac{1}{4M}\cdot \frac{(M-1)(M-2)}{2}|x_1|^2\\
		&=\left(\frac{5}{4}M-\frac{7}{4}+\frac{1}{2M}\right)|x_1|^2
		\end{align*}
		which implies
		\[
		\|\ma_{\Omega^c\backslash\{1\}}\vx_{\Omega^c\backslash\{1\}}\|_2\le \sqrt{\frac{5}{4}M-\frac{7}{4}+\frac{1}{2M}}|x_1|.
		\]
		\item For $\|\ma_{\Omega}\ma_{\Omega}^{\dag}(\ma_{\Omega^c}\vx_{\Omega^c})\|_2$:
		
		We have that 
		\[
		\begin{aligned}
		\|\ma_{\Omega}\ma_{\Omega}^{\dag}(\ma_{\Omega^c}\vx_{\Omega^c})\|_2^2&=|\la\ma_{\Omega}^*\ma_{\Omega^c}\vx_{\Omega^c},(\ma_{\Omega}^*\ma_{\Omega})^{-1}\ma_{\Omega}^*\ma_{\Omega^c}\vx_{\Omega^c} \ra|\\
		&\le\|\ma_{\Omega}^*\ma_{\Omega^c}\vx_{\Omega^c}\|_2\|(\ma_{\Omega}^*\ma_{\Omega})^{-1}\|\|\ma_{\Omega}^*\ma_{\Omega^c}\vx_{\Omega^c}\|_2\\
		&\le \frac{\sqrt{M}|x_1|}{4}\cdot \frac{4}{3}\cdot\frac{\sqrt{M}|x_1|}{4}\quad  (\text{from }\eqref{eq:estimate})\\
		&\le \frac{M|x_1|^2}{12}    .    \end{aligned}
		\]
		\item For $\|\ee-\ma_{\Omega}\ma_{\Omega}^{\dag}\ee\|_2$:
		
		Since $\ee-\ma_{\Omega}\ma_{\Omega}^{\dag}\ee$ is an orthogonal projection of $\ee$, we have that
		\[
		\|\ee-\ma_{\Omega}\ma_{\Omega}^{\dag}\ee\|_2\le \|\ee\|_2.
		\]
	\end{itemize}
	
	Overall, in order to show \eqref{eq:noiseless}, it suffices to show that
	\[
	(1-\tau^2)\left(\frac{2|x_1|}{3}-\frac{4}{3}\tau\|\ee\|_2\right)^2>\tau^2\left[\left(\sqrt{\frac{5}{4}M-\frac{7}{4}+\frac{1}{2M}}+\frac{1}{\sqrt{12}}\sqrt{M}\right)|x_1|+\|\ee\|_2\right]^2,
	\]
	or equivalently,
	\begin{equation}\label{eq:noisestrength}
	\underbrace {\left[ {\frac{2}{3}\sqrt {\frac{{1 - {\tau ^2}}}{{{\tau ^2}}}}  - \left( {\sqrt {\frac{5}{4}M - \frac{7}{4} + \frac{1}{{2M}}}  + \frac{1}{{\sqrt {12} }}\sqrt M } \right)} \right]}_F|{x_1}| > \underbrace {\left(\frac{4}{3}\sqrt {1 - {\tau ^2}}  + 1\right)}_G\|\ee\|_2.
	\end{equation}
	We choose
	\begin{equation}
	f(M,\tau)=\frac{F}{G}. \label{eq:functionf}
	\end{equation}
	With $\tau \le 1/(\sqrt{6M})$ (implied from \eqref{threspara} and \eqref{mutualcondition}), we note that $F>0$ for $M\ge 1$.  We then have if $\|\ee\|_2 \le f(M,\tau)\min_{i\in \supp(\vx)}(|\vx_i|)$, then the inequality \eqref{eq:noisestrength} holds. Consequently, the inequality \eqref{eq:noiseless} holds, and thus \eqref{supportineq} holds. As noted from the beginning of the proof, the above deterministic analysis is done conditioned on the event $\mathcal O$, which holds with probability $1-2/N^{\kappa}$. The proof is complete.
\end{proof}

\subsection{Discussion}
In this paper, we consider a sparse recovery algorithm named Thresholding Greedy Pursuit (TGP). The algorithm is based on CoSaMP with the addition of thresholding procedure. The only assumption we need for the probability distribution of the noise is rotational invariance. No knowledge about its strength is needed. This assumption is reasonable in high dimension, and so we can borrow techniques from high dimensional probability. 

\begin{figure*}[ht] \centering 
	\captionsetup{width=0.8\linewidth}
	\subfloat{\includegraphics[width=0.333\textwidth]{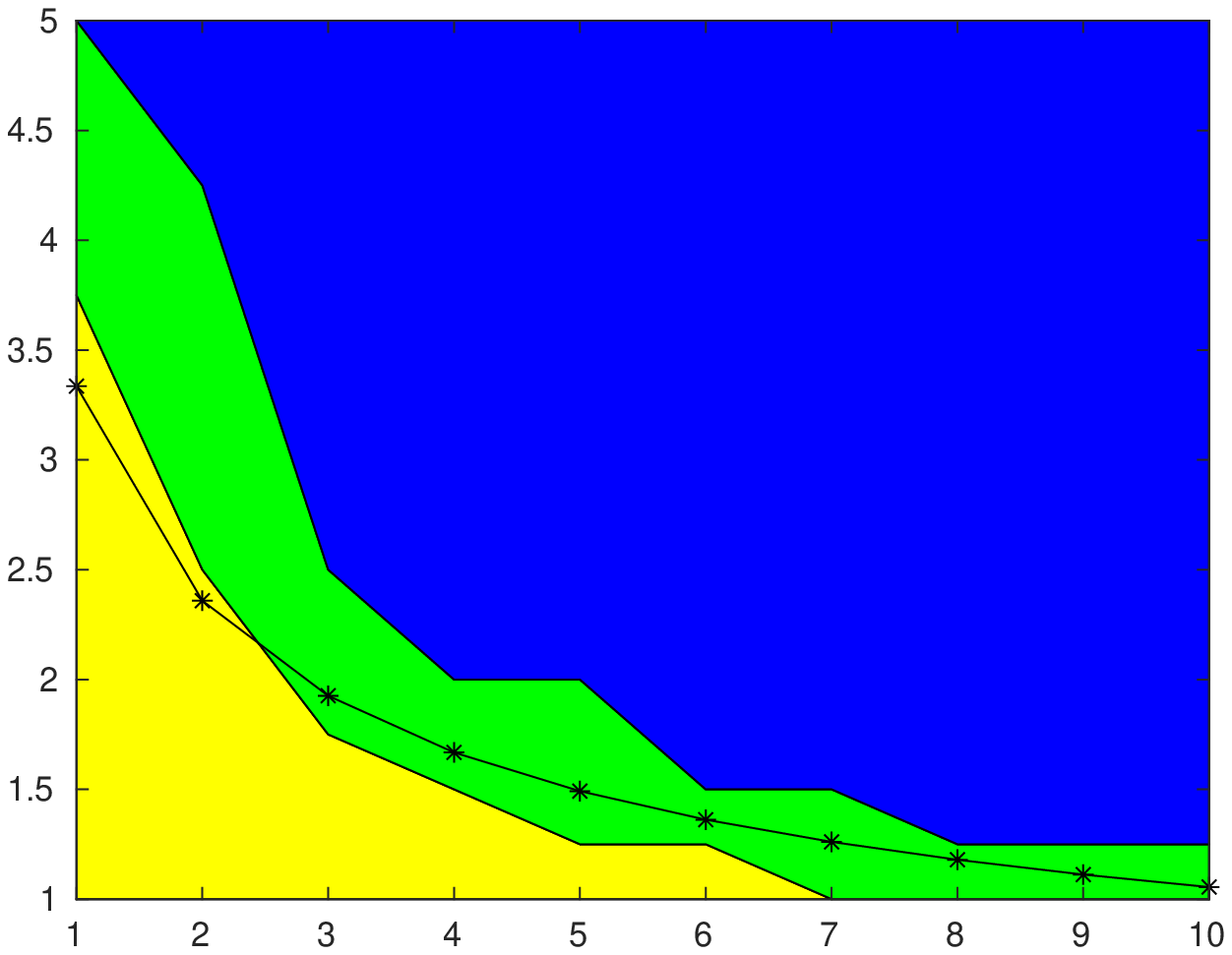}}  \subfloat{\includegraphics[width=0.333\textwidth]{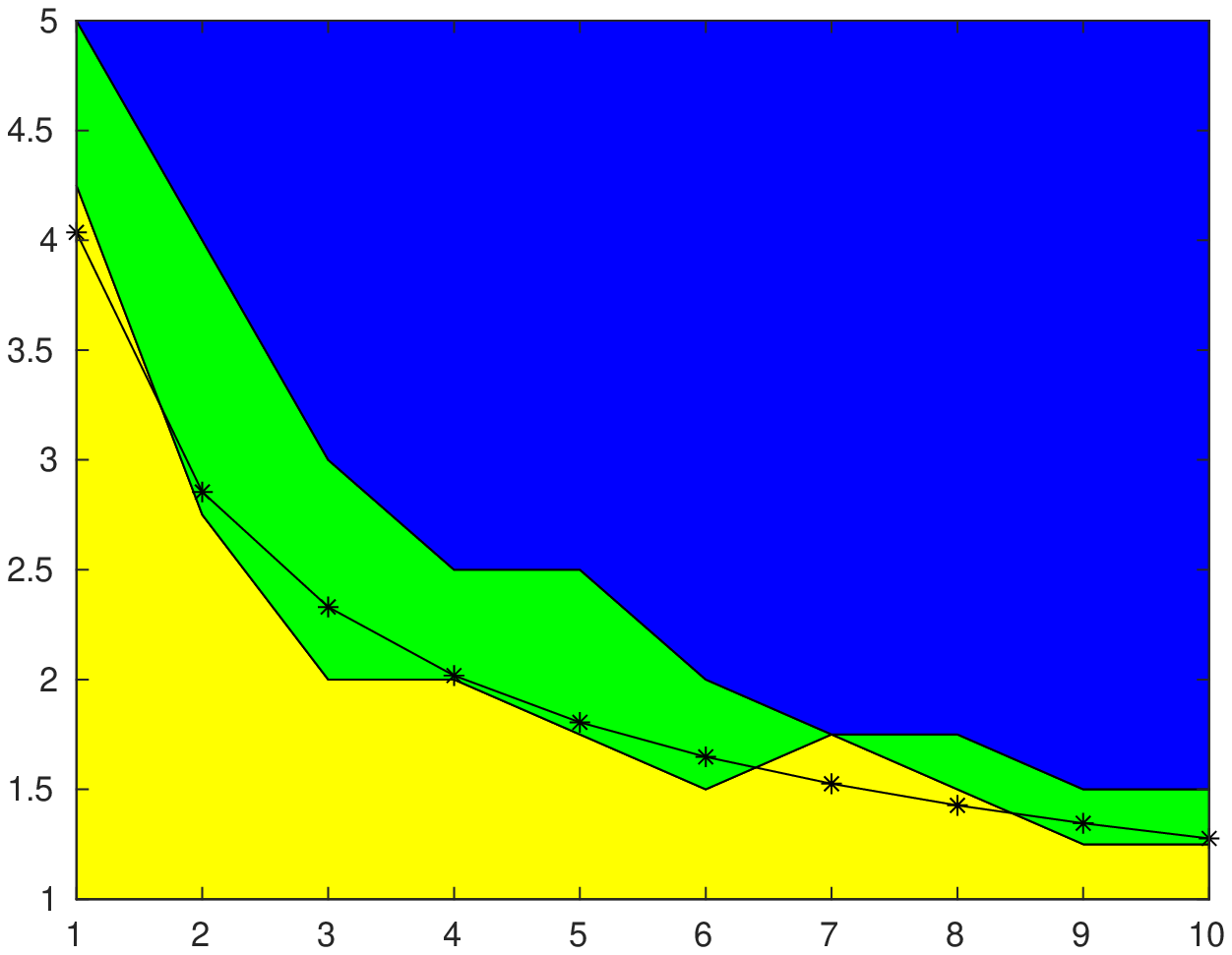}}
	\subfloat{\includegraphics[width=0.333\textwidth]{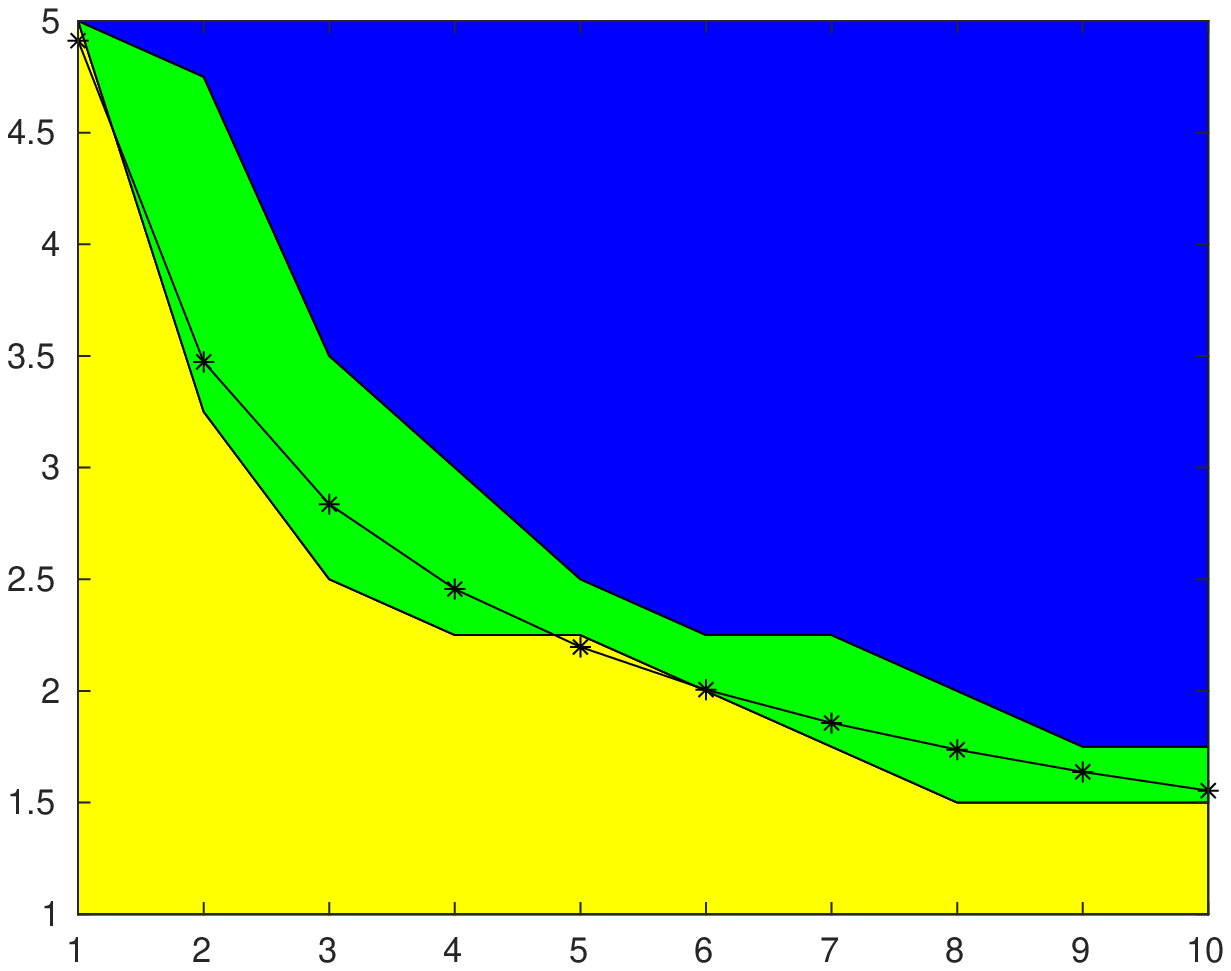}}
	\caption{Algorithm performance for exact support recovery. Success corresponds to a value of one (yellow), and failure corresponds to a value of zero (blue). The small phase transition zone (green) contains intermediate values. The black lines are the theoretical estimate $\sqrt{N}/\sqrt{M\log N}$. Ordinate and abscissa are the sparsity $M$ and $\|\ee\|_2/\|\bb_0\|_2$. The data sizes are $N=400$ (left), $N=630$ (center), and $N=1000$ (right).}
	\label{performance}
\end{figure*}

We analyze the performance of TGP algorithm in the regime where $N$ tends to infinity. As mentioned in the proofs, the sparsity level $M$ is required to be of level less than $O(\sqrt{N}/\sqrt{\log N})$. For bigger $M$, the number of measurements $N$ needs to increase as well. In Figure \ref{performance}, we illustrate the performance of TGP for different sparsity levels $M$ and $\|\ee\|_2/\|\bb_0\|_2$. In this experiment, the matrix $\ma$ consists of normally distributed columns, and the signal $\vx$ is uniformly 1 at nonzero locations. Success in recovering the true support of the unknown corresponds to a value of one (the color yellow), and failure corresponds to a value of zero (the color blue). The small phase transition zone (the color green) contains intermediate values. The black lines are the graphs of $\sqrt{N}/\sqrt{M\log N}$ which represent the relative level of noise $\|\ee\|_2/\|\bbo\|_2$ the algorithm can sustain. This can be seen from \eqref{eq:functionf} when $x_1=1$ that $\|\ee\|_2\lesssim \sqrt{N}/\sqrt{\log N}$ since $\tau \approx \sqrt{\log N}/\sqrt{N}$.

We also tested TGP for the case when the measurement matrix has some of the columns that are close to collinear. In general, if a location in the support corresponds to one of those columns, TGP will return all of the columns that are nearby. This is reasonable as the corresponding dot products will be approximately equal and the Thresholding procedure will identify all of them. This amplifies the nature of the algorithm: the more we know about the measurement matrix $\ma$, the better we can design the thresholding parameter $\tau$. A possible research in this direction is to ensure no false negatives present when there are clusters of columns that are close to each other.

The MATLAB codes for the TGP and for the comparison with CoSaMP are available at \url{https://github.com/randomwalk94/TGP}.

{\textbf{Acknowledgement.} We are grateful to Anna Gilbert for suggesting to apply the conjugate gradient approach of CoSaMP to design of an iterative algorithm to solve Square-Root LASSO.
	This work was partially supported by NSF DMS-1813943 and AFOSR FA9550-20-1-0026.}

\vskip 0.2in
\bibliographystyle{alpha}
\bibliography{tgp}

\newcommand{\etalchar}[1]{$^{#1}$}
\begin{thebibliography}{MNPT20b}

\bibitem[BCW11]{BELLONI2011}
A.~Belloni, V.~Chernozhukov, and L.~Wang.
\newblock {Square-root LASSO: pivotal recovery of sparse signals via conic
  programming}.
\newblock {\em Biometrika}, 98:791--806, 2011.

\bibitem[BDF{\etalchar{+}}11]{Bourgain2011}
Jean Bourgain, Stephen Dilworth, Kevin Ford, Sergei Konyagin, and Denka
  Kutzarova.
\newblock {Explicit constructions of rip matrices and related problems}.
\newblock {\em Duke Mathematical Journal}, 159(1):145--185, 2011.

\bibitem[CD94]{Chen1994}
Shaobing Chen and David Donoho.
\newblock {Basis Pursuit}.
\newblock In {\em Proceedings of Asilomar Conference on Signals, Systems and
  Computers}, volume~1, pages 41--44. IEEE Computer Society, 1994.

\bibitem[CDS01]{Chen2001}
S.~S. Chen, D.~L. Donoho, and M.~A. Saunders.
\newblock {Atomic decomposition by basis pursuit}.
\newblock {\em SIAM Review}, 43(1):129--159, 2001.

\bibitem[CLWB16]{Chichignoud2016}
Micha{\"{e}}l Chichignoud, Johannes Lederer, Martin~J Wainwright, and Francis
  Bach.
\newblock {A Practical scheme and fast algorithm to tune the LASSO with
  optimality guarantees}.
\newblock {\em Journal of Machine Learning Research}, 17:1--17, 2016.

\bibitem[CM73]{Claerbout1973}
Jon~F. Claerbout and Francis Muir.
\newblock {Robust modeling with erratic data}.
\newblock {\em Geophysics}, 38(5):826--844, 1973.

\bibitem[CMN{\etalchar{+}}19]{CrespoMarques2019}
E.~{Crespo Marques}, N.~{Maciel}, L.~{Naviner}, H.~{Cai}, and J.~{Yang}.
\newblock A review of sparse recovery algorithms.
\newblock {\em IEEE Access}, 7:1300--1322, 2019.

\bibitem[CRT06a]{Candes2006}
Emmanuel~J. Cand{\`{e}}s, Justin Romberg, and Terence Tao.
\newblock {Robust uncertainty principles: Exact signal reconstruction from
  highly incomplete frequency information}.
\newblock {\em IEEE Transactions on Information Theory}, 52(2):489--509, 2006.

\bibitem[CRT06b]{Candes2006a}
Emmanuel~J. Cand{\`{e}}s, Justin~K. Romberg, and Terence Tao.
\newblock {Stable signal recovery from incomplete and inaccurate measurements}.
\newblock {\em Communications on Pure and Applied Mathematics},
  59(8):1207--1223, 2006.

\bibitem[DDD04]{Daubechies2004}
I.~Daubechies, M.~Defrise, and C.~{De Mol}.
\newblock {An iterative thresholding algorithm for linear inverse problems with
  a sparsity constraint}.
\newblock {\em Communications on Pure and Applied Mathematics},
  57(11):1413--1457, 2004.

\bibitem[DDPS20]{Dileep2020}
B.~P.~V. Dileep, Pranab~K. Dutta, P.~M.K. Prasad, and M.~Santhosh.
\newblock {Sparse recovery based compressive sensing algorithms for diffuse
  optical tomography}.
\newblock {\em Optics and Laser Technology}, 128:106234, 2020.

\bibitem[DTDS12]{Donoho2012}
David~L. Donoho, Yaakov Tsaig, Iddo Drori, and Jean~Luc Starck.
\newblock {Sparse solution of underdetermined systems of linear equations by
  stagewise orthogonal matching pursuit}.
\newblock {\em IEEE Transactions on Information Theory}, 58(2):1094--1121,
  2012.

\bibitem[EB02]{Elad2002}
Michael Elad and Alfred~M. Bruckstein.
\newblock {A generalized uncertainty principle and sparse representation in
  pairs of bases}.
\newblock {\em IEEE Transactions on Information Theory}, 48(9):2558--2567,
  2002.

\bibitem[EK09]{Eldar2009}
Yonina~C. Eldar and Gitta Kutyniok.
\newblock {\em {Compressed sensing: Theory and applications}}.
\newblock Cambridge University Press, 2009.

\bibitem[FN03]{Feuer2003}
Arie Feuer and Arkadi Nemirovski.
\newblock {On sparse representation in pairs of bases}.
\newblock {\em IEEE Transactions on Information Theory}, 49(6):1579--1581,
  2003.

\bibitem[FNW07]{Figueiredo2007}
M{\'{a}}rio~A.T. Figueiredo, Robert~D. Nowak, and Stephen~J. Wright.
\newblock {Gradient projection for sparse reconstruction: Application to
  compressed sensing and other inverse problems}.
\newblock {\em IEEE Journal on Selected Topics in Signal Processing},
  1(4):586--597, 2007.

\bibitem[Fuc05]{Fuchs2005}
Jean~Jacques Fuchs.
\newblock {Recovery of exact sparse representations in the presence of bounded
  noise}.
\newblock {\em IEEE Transactions on Information Theory}, 51(10):3601--3608,
  2005.

\bibitem[GBK16]{Gupta2016}
Manish Gupta, Scott~A. Beckett, and Elizabeth~B. Klerman.
\newblock {On-line EEG denoising using correlated sparse recovery}.
\newblock In {\em International Symposium on Medical Information and
  Communication Technology, ISMICT}, volume 2016-June. IEEE Computer Society,
  2016.

\bibitem[GGI{\etalchar{+}}02a]{Gilbert2002}
A.~C. Gilbert, S.~Guha, P.~Indyk, S.~Muthukrishnan, and M.~Strauss.
\newblock {Near-optimal sparse fourier representations via sampling}.
\newblock In {\em Proceedings of the thiry-fourth annual ACM symposium on
  Theory of computing - STOC '02}, page 152, New York, New York, USA, 2002. ACM
  Press.

\bibitem[GGI{\etalchar{+}}02b]{Gilbert2002a}
Anna~C. Gilbert, Sudipto Guha, Piotr Indyk, Yannis Kotidis, S.~Muthukrishnan,
  and Martin~J. Strauss.
\newblock {Fast, small-space algorithms for approximate histogram maintenance}.
\newblock In {\em Proceedings of the thiry-fourth annual ACM symposium on
  Theory of computing - STOC '02}, page 389, New York, New York, USA, 2002. ACM
  Press.

\bibitem[GGMS03]{Gilbert2003}
A.~C. Gilbert, A.~C. Gilbert, S.~Muthukrishnan, and M.~Strauss.
\newblock {Improved time bounds for near-optimal sparse fourier
  representations}.
\newblock {\em Proceedings of SPIE, Volume 5914 Wavelets XI}, 2003.

\bibitem[GGS{\etalchar{+}}06]{Gilbert2006}
A.~C. Gilbert, A.~C. Gilbert, M.~J. Strauss, J.~A. Tropp, and R.~Vershynin.
\newblock {Algorithmic linear dimension reduction in the l1 norm for sparse
  vectors}.
\newblock {\em Allerton 2006 (44TH Annual Allerton Conference on Communication,
  Control, and Computing)}, 2006.

\bibitem[GSTV07]{Gilbert2007}
A.~C. Gilbert, M.~J. Strauss, J.~A. Tropp, and R.~Vershynin.
\newblock {One sketch for all: Fast algorithms for compressed sensing}.
\newblock In {\em Proceedings of the Annual ACM Symposium on Theory of
  Computing}, pages 237--246, New York, New York, USA, 2007. ACM Press.

\bibitem[GV12]{Golub2012}
Gene Golub and Charles {Van Loan}.
\newblock {\em {Matrix Computations}}.
\newblock Johns Hopkins University Press, 2012.

\bibitem[HM18]{Homrighausen2018}
Darren Homrighausen and Daniel~J. McDonald.
\newblock {A study on tuning parameter selection for the high-dimensional
  lasso}.
\newblock {\em Journal of Statistical Computation and Simulation},
  88(15):2865--2892, 2018.

\bibitem[KJ18]{Kueng2018}
Richard Kueng and Peter Jung.
\newblock {Robust nonnegative sparse recovery and the nullspace property of 0/1
  measurements}.
\newblock {\em IEEE Transactions on Information Theory}, 64(2):689--703, 2018.

\bibitem[LDB09]{Laska2009}
Jason~N. Laska, Mark~A. Davenport, and Richard~G. Baraniuk.
\newblock {Exact signal recovery from sparsely corrupted measurements through
  the pursuit of justice}.
\newblock In {\em Conference Record - Asilomar Conference on Signals, Systems
  and Computers}, pages 1556--1560, 2009.

\bibitem[MNPT20a]{Moscoso2020}
Miguel Moscoso, Alexei Novikov, George Papanicolaou, and Chrysoula Tsogka.
\newblock {Imaging with highly incomplete and corrupted data}.
\newblock {\em Inverse Problems}, 36(3):035010, 2020.

\bibitem[MNPT20b]{Moscoso2020a}
Miguel Moscoso, Alexei Novikov, George Papanicolaou, and Chrysoula Tsogka.
\newblock {The Noise Collector for sparse recovery in high dimensions}.
\newblock {\em PNAS}, 117(21):11226--11232, 2020.

\bibitem[MZ93]{Mallat1993}
Stephane~G. Mallat and Zhifeng Zhang.
\newblock {Matching pursuits with time-frequency dictionaries}.
\newblock {\em IEEE Transactions on Signal Processing}, 41(12):3397--3415,
  1993.

\bibitem[NT09]{Needell2009}
D.~Needell and J.~A. Tropp.
\newblock {CoSaMP: Iterative signal recovery from incomplete and inaccurate
  samples}.
\newblock {\em Applied and Computational Harmonic Analysis}, 26(3):301--321,
  2009.

\bibitem[NV09]{Needell2009a}
Deanna Needell and Roman Vershynin.
\newblock {Uniform uncertainty principle and signal recovery via regularized
  orthogonal matching pursuit}.
\newblock {\em Foundations of Computational Mathematics}, 9(3):317--334, 2009.

\bibitem[NV10]{Needell2010}
Deanna Needell and Roman Vershynin.
\newblock {Signal recovery from incomplete and inaccurate measurements via
  regularized orthogonal matching pursuit}.
\newblock {\em IEEE Journal on Selected Topics in Signal Processing},
  4(2):310--316, 2010.

\bibitem[TG07]{Tropp2007}
Joel~A. Tropp and Anna~C. Gilbert.
\newblock {Signal recovery from random measurements via orthogonal matching
  pursuit}.
\newblock {\em IEEE Transactions on Information Theory}, 53(12):4655--4666,
  2007.

\bibitem[Tib96]{Tibshirani1996}
Robert Tibshirani.
\newblock {Regression shrinkage and selection via the LASSO}.
\newblock {\em Journal of the Royal Statistical Society: Series B
  (Methodological)}, 58(1):267--288, 1996.

\bibitem[Tro06]{Tropp2006}
Joel~A. Tropp.
\newblock {Just relax: Convex programming methods for identifying sparse
  signals in noise}.
\newblock {\em IEEE Transactions on Information Theory}, 52(3):1030--1051,
  2006.

\bibitem[Ver18]{Vershynin2018}
Roman Vershynin.
\newblock {\em {High-Dimensional Probability}}.
\newblock Cambridge University Press, 2018.

\bibitem[Wai09]{Wainwright2009}
Martin~J. Wainwright.
\newblock {Sharp thresholds for high-dimensional and noisy sparsity recovery
  using l1-constrained quadratic programming (Lasso)}.
\newblock {\em IEEE Transactions on Information Theory}, 55(5):2183--2202,
  2009.

\bibitem[Yan13]{Yang}
Jie Yang.
\newblock {\em {A machine learning paradigm based on sparse signal
  representation}}.
\newblock PhD thesis, University of Wollongong, 2013.

\bibitem[YD15]{Yang2015a}
Mingrui Yang and Frank {De Hoog}.
\newblock {Orthogonal matching pursuit with thresholding and its application in
  compressive sensing}.
\newblock {\em IEEE Transactions on Signal Processing}, 63(20):5479--5486,
  2015.

\bibitem[Zou06]{Zou2006}
Hui Zou.
\newblock {The Adaptive LASSO and its oracle properties}.
\newblock {\em Journal of the American Statistical Association},
  101(476):1418--1429, 2006.

\end{thebibliography}

\end{document}